\documentclass[conference]{IEEEtran}

\usepackage[noadjust]{cite}
\usepackage{ifthen}
\newboolean{blind}   
\setboolean{blind}{false}   

\usepackage{xcolor}
\usepackage{colortbl}
\usepackage{booktabs}
\usepackage{graphicx}
\usepackage{multirow}
\usepackage{pgfplots}
\usepackage{ifthen}
\usepackage{subcaption}
\usepackage{amsmath,amsthm,amssymb,amsxtra} 
\usepackage{mathtools}
\usepackage{float} 

\usepgfplotslibrary{colormaps}
\pgfplotsset{compat=newest} 
\pgfdeclarelayer{background}
\pgfdeclarelayer{background_1}
\pgfdeclarelayer{background_2}
\pgfdeclarelayer{background_3}
\pgfdeclarelayer{foreground_6}
\pgfdeclarelayer{foreground_5}
\pgfdeclarelayer{foreground_4}
\pgfdeclarelayer{foreground_3}
\pgfdeclarelayer{foreground_2}
\pgfdeclarelayer{foreground_1}
\pgfdeclarelayer{foreground}
\pgfsetlayers{background,background_1,background_2,background_3,main,foreground_6,foreground_5,foreground_4,foreground_3,foreground_2,foreground_1,foreground}

\usepackage{tikz}
\usetikzlibrary{arrows, arrows.meta, positioning, quotes, shapes, fit, calc, decorations.pathreplacing, calligraphy, plotmarks, external}
\tikzstyle{block} = [draw, thick, text=black, align=center, minimum width=1.2cm, minimum height=0.6cm, font=\small]
\tikzstyle{Block} = [draw, thick, rounded corners, text=black, align=center, minimum width=2cm, minimum height=1.2cm, font=\small]
\usepackage{circuitikz}
\ctikzset{bipoles/cuteswitch/thickness=0.3}

\usepackage{amsmath}
\usepackage[binary-units=true]{siunitx}
\DeclareSIUnit\BL{\giga\bit\kilo\meter\per\second}
\DeclareSIUnit\FL{\decibel\per\kilo\meter}
\DeclareSIUnit\CD{\pico\second\per\nano\meter\per\kilo\meter}

\usepackage{mleftright}
\usepackage{relsize}
\usepackage{svg}
\usepackage{textgreek}
\usepackage{bm}
\usepackage{verbatim}
\usepackage{tabularray}
\usepackage{makecell}

\usepackage[nohyperlinks,nolist]{acronym}

\acrodef{ai}[AI]{artificial intelligence}
\acrodef{ask}[ASK]{amplitude-shift keying}
\acrodef{awgn}[AWGN]{additive white Gaussian noise}
\acrodef{ann}[ANN]{artificial neural network}
\acrodef{asic}[ASIC]{application-specific integrated circuit}

\acrodef{bce}[BCE]{binary cross-entropy}
\acrodef{ber}[BER]{bit error rate}
\acrodef{bler}[BLER]{block error rate}
\acrodef{bpsk}[BPSK]{binary phase-shift keying}
\acrodef{bram}[BRAM]{block random access memory}
\acrodef{bilstm}[biLSTM]{bidirectional long short-term memory}
\acrodef{bp}[BP]{backward pass}

\acrodef{cd}[CD]{chromatic dispersion}
\acrodef{cir}[CIR]{combined impulse response}
\acrodef{cma}[CMA]{constant modulus algorithm}
\acrodef{cnn}[CNN]{convolutional neural network}
\acrodef{cpu}[CPU]{central processing unit}
\acrodef{ctp}[CTP]{channel transition probability}
\acrodefplural{ctp}[CTPs]{channel transition probabilities}
\acrodef{csi}[CSI]{channel state information}
\acrodef{conv1d}[conv1d]{one-dimensional convolution}

\acrodef{dnn}[DNN]{deep neural network}
\acrodef{dsp}[DSP]{digital signal processor}
\acrodef{dram}[DRAM]{dynamic random access memory}
\acrodef{dfe}[DFE]{decision-feedback equalizer}
\acrodef{dop}[DOP]{degree of paralellism}
\acrodef{dbp}[DBP]{digital back-propagation}

\acrodef{fc}[FC]{fully connected}
\acrodef{fec}[FEC]{forward error correction}
\acrodef{fir}[FIR]{finite impulse response}
\acrodef{fifo}[FIFO]{first in first out}
\acrodef{fpga}[FPGA]{field programmable gate array}
\acrodef{fp}[FP]{forward pass}
\acrodef{ff}[FF]{flip-flop}

\acrodef{elu}[ELU]{exponential linear unit}

\acrodef{gan}[GAN]{generative adversarial network}
\acrodef{gpu}[GPU]{graphics processing unit}
\acrodef{gops}[GOPS]{giga operations per second}

\acrodef{hls}[HLS]{high-level synthesis}
\acrodef{hwa}[HWA]{historical weight averaging}

\acrodef{imdd}[IM/DD]{intensity modulation with direct detection}
\acrodef{iot}[IoT]{Internet of things}
\acrodef{isi}[ISI]{inter-symbol interference}

\acrodef{ldpc}[LDPC]{low-density parity-check}
\acrodef{lms}[LMS]{least-mean-square}
\acrodef{lut}[LUT]{look-up table}

\acrodef{mlsd}[MLSD]{maximum-likelihood sequence detection}
\acrodef{mse}[MSE]{mean-squared-error}
\acrodef{mlse}[MLSE]{maximum likelihood sequence estimator}
\acrodef{mac}[MAC]{multiply-accumulate}
\acrodef{ma}[MA]{moving average}
\acrodef{map}[MAP]{maximum a posteriori}
\acrodef{msm}[MSM]{merge stream module}
\acrodef{mzm}[MZM]{mach-zehnder modulator}
\acrodef{nn}[NN]{neural network}
\acrodef{ns}[NS]{non-saturating}

\acrodef{ogm}[OGM]{overlap generate module}
\acrodef{orm}[ORM]{overlap remove module}
\acrodef{onu}[ONU]{optical network unit}
\acrodef{olt}[OLT]{optical line terminal}
\acrodef{odn}[ODN]{optical distribution network}

\acrodef{pam}[PAM]{pulse-amplitude modulation}
\acrodef{pdf}[pdf]{probability density function}
\acrodef{pe}[PE]{processing element}
\acrodef{pmf}[pmf]{probability mass function}
\acrodef{pon}[PON]{passive optical network}

\acrodef{qlf}[QLF]{quantization loss factor}
\acrodef{qam}[QAM]{quadrature amplitude modulation}

\acrodef{rc}[RC]{raised-cosine}
\acrodef{relu}[ReLU]{rectified linear unit}
\acrodef{rls}[RLS]{recursive least squares}
\acrodef{rnn}[RNN]{recurrent neural network}

\acrodef{ser}[SER]{symbol error rate}
\acrodef{sgd}[SGD]{stochastic gradient descent}
\acrodef{sld}[SLD]{square-law detection}
\acrodef{snr}[SNR]{signal-to-noise ratio}
\acrodef{sps}[sps]{samples per symbol}
\acrodef{ssmf}[SSMF]{standard single-mode fiber}
\acrodef{simd}[SIMD]{single instruction multiple data}
\acrodef{ssm}[SSM]{split stream module}
\acrodef{spb}[SPB]{symbols per batch}
\acrodef{slda}[SLDA]{streaming linear discriminant analysis}

\acrodef{ti}[TI]{training iteration}

\acrodef{vnle}[VNLE]{Volterra-based nonlinear equalizer}

\newcommand{\todo}[1]{\textcolor{red}{\textbf{[ToDo: #1]}}}

\newcommand{\mlr}[1]{\mleft(#1\mright)}

\definecolor{RPTU_BlueGray}{RGB}{80,114,137}
\definecolor{RPTU_GreenGray}{RGB}{119,182,186}
\definecolor{RPTU_DarkBlue}{RGB}{4,44,88}
\definecolor{RPTU_LightBlue}{RGB}{106,178,231}
\definecolor{RPTU_DarkGreen}{RGB}{0,107,107}
\definecolor{RPTU_LightGreen}{RGB}{38,208,124}
\definecolor{RPTU_Violett}{RGB}{76,53,117}
\definecolor{RPTU_Pink}{RGB}{209,56,150}
\definecolor{RPTU_Red}{RGB}{227,27,76}
\definecolor{RPTU_Orange}{RGB}{255,162,82}
\definecolor{RPTU_Black}{RGB}{0,0,0}
\definecolor{RPTU_White}{RGB}{255,255,255}

\AtBeginDocument{%
  \providecommand\BibTeX{{%
    \normalfont B\kern-0.5em{\scshape i\kern-0.25em b}\kern-0.8em\TeX}}}

\begin{document}

\title{Achieving High Throughput with a Trainable Neural-Network-Based Equalizer for Communications on FPGA}

\ifthenelse{\boolean{blind}}
{
\author{\huge \textit{Authors omitted due to double-blind review}}
}
{

\author{
\IEEEauthorblockN{Jonas Ney and Norbert Wehn}
\IEEEauthorblockA{\textit{Microelectronic Systems Design (EMS)}, \textit{RPTU Kaiserslautern-Landau}, Germany \\
\{\texttt{jonas.ney}, \texttt{norbert.wehn}\}\texttt{@rptu.de}} 
}
}

\maketitle

\begin{abstract}
The ever-increasing data rates of modern communication systems lead to severe distortions of the communication signal, imposing great challenges to state-of-the-art signal processing algorithms. In this context, \ac{nn}-based equalizers are a promising concept since they can compensate for impairments introduced by the channel. However, due to the large computational complexity, efficient hardware implementation of \acp{nn} is challenging. Especially the backpropagation algorithm, required to adapt the \ac{nn}'s parameters to varying channel conditions, is highly complex, limiting the throughput on resource-constrained devices like \acp{fpga}. In this work, we present an \ac{fpga} architecture of an \ac{nn}-based equalizer that exploits batch-level parallelism of the convolutional layer to enable a custom mapping scheme of two multiplication to a single \ac{dsp}. Our implementation achieves a throughput of up to 20 GBd, which enables the equalization of high-data-rate nonlinear optical fiber channels while providing adaptation capabilities by retraining the \ac{nn} using backpropagation. As a result, our \ac{fpga} implementation outperforms an embedded \ac{gpu} in terms of throughput by two orders of magnitude. Further, we achieve a higher energy efficiency and throughput as state-of-the-art  \ac{nn} training \ac{fpga} implementations. Thus, this work fills the gap of high-throughput \ac{nn}-based equalization while enabling adaptability by \ac{nn} training on the edge \ac{fpga}. 
\end{abstract}

\ifthenelse{\boolean{blind}}
{
\let\thefootnote\relax\footnotetext{\centering \LARGE \textit{Grant agreements omitted due to double-blind review}}
}
{
\let\thefootnote\relax\footnotetext{This work was funded by the German Federal Ministry of Education and Research (BMBF) under grant agreements 16KIS1316 (AI-NET-ANTILLAS) and 16KISK004 (Open6GHuB). Further it was funded by the Carl Zeiss Stiftung under the Sustainable Embedded AI project (P2021-02-009).}
}

\section{Introduction}

In recent years, there has been a remarkable boost in the throughput of communication systems, driven by the rising demand for high-speed data transmission in various applications such as data centers, video streaming, and cloud computing \cite{Bayvel2016}. As a result, peak data rates have increased to \SI{20}{\giga bit \per \second} for 5G \cite{Kaur2020} and are expected to reach \SI{1000}{\giga bit \per \second} for the upcoming 6G standard  \cite{Rajatheva2020}. 
Those immense advancements in throughput and data rate come at the cost of increased distortions of the communication signal due to nonlinearities and \ac{isi} e.g. caused by multipath propagation for wireless communication systems or \ac{cd} for optical fiber communications. To compensate for those distortions and to achieve reliable communication, more advanced signal-processing algorithms are required. 

A promising candidate to tackle the challenges of modern high-throughput communication systems are \ac{nn}-based algorithms \cite{zerguine2001, schaedler2019, Ney2022}. Especially the equalizer, responsible for mitigating the impairments of the communication channel to increase the reliability of the system, is a component that can benefit from the latest advancements of \ac{nn} research. In particular, \acp{nn} have shown remarkable results for channels with nonlinear effects (e.g. high-data-rate optical fiber channels), for which no exact analytical solutions exist for equalization~\cite{Khan2019, Ney2023_2}. 

Besides throughput, key features of next-gen communication systems include flexibility and adaptability \cite{Yazar2020}. Those aspects are especially critical in the context of changing environmental conditions and transmission characteristics. In general, \ac{nn}-based systems are well suited to fulfill the strict adaptability requirements. By retraining or fine-tuning the \ac{nn}, it can adapt to the changing conditions by updating its trainable parameters \cite{Ney2023}. However, updating the weights and bias of the \ac{nn} requires complex optimization algorithms like backpropagation and gradient descent. Performing these algorithms in real time while meeting the high-throughput constraints is challenging on conventional platforms.

In contrast, \acp{fpga} are a promising platform to satisfy the strict throughput and flexibility requirements since they provide a huge parallelism and customizability. For certain low-complex \acp{nn}, which are commonly applied in communications, they can even outperform high-performance \acp{gpu} in terms of throughput~\cite{Ney2023_2}. 
Additionally, an \ac{fpga} design is a first step towards a custom \ac{asic} as used in practical communication systems. 
Thus, in this work, we present a trainable \ac{fpga} implementation of a \ac{cnn}-based equalizer that satisfies the high-throughput requirements of an optical communication channel while providing retraining capabilities to enable the required adaptability.  

In summary, our novel contributions are the following: 
\begin{itemize}
    \item A high-throughput \ac{fpga} implementation of a \ac{cnn}-based equalizer \textit{including} backpropagation featuring an adjustable parallelism for inference and training. 
    \item An optimized \ac{conv1d} layer architecture featuring a custom mapping scheme of two multiplications to a single \ac{dsp} to increase the achievable throughput on resource-constrained \acp{fpga}.
    \item An in-depth analysis of the convergence behavior of our \ac{cnn}-based equalizer for a \SI{20}{\giga Bd} optical fiber channel. 
    \item A detailed comparison of our approach to an embedded \ac{gpu} implementation and to other state-of-the-art \ac{fpga} \ac{nn} training implementations
\end{itemize}

\section{Related Work}
\label{sec:related work}

In the following, we discuss the most relevant works related to our high-throughput, retrainable \ac{fpga} implementation of an \ac{nn}-based equalizer. First, the state-of-the-art in the field of \ac{nn} training on \ac{fpga} is presented, and afterwards, \ac{fpga} implementations of \ac{nn}-based equalizers are introduced.    

\subsection{\ac{nn} Training on \ac{fpga}}
\label{sec:nn_training_on_fpga}

In recent years research strongly focused on the acceleration of \ac{nn} inference using \acp{fpga}~\cite{Li2018, Duan2018, Ney2023_2}. In contrast, for the computationally intensive \ac{nn} training, \acp{gpu} still remain the platform of choice for the majority of applications. Only a few works tackle the challenges of implementing the \ac{nn} training including backpropagation and weight update on \ac{fpga}. 

An automatic compiler-based \ac{fpga} accelerator for \ac{cnn} training was presented in~\cite{Venkataramanaiah2019}. It automatically generates an \ac{fpga}-synthesizable hardware description including a novel cyclic weight storage access scheme. As a result, a throughput of up to \SI{479}{GOPS} is reported. 
In 2017, Liu et al. proposed an \ac{fpga}-based processor for training \acp{cnn}~\cite{Liu2017}. They designed a uniform computation engine design and a scalable framework that exploits parallelism to outperform an Intel i5 \ac{cpu} by \num{10} times regarding processing time. 
Mazouz et al. proposed an automated \ac{cnn} back-propagation pipeline generation for \ac{fpga} online training in 2021~\cite{Mazouz2021}. A speed-up of \num{5.8} times was reported as compared to the Nvidia GTX 1050 Ti \ac{gpu}.
An approach to accelerate continual learning was targeted by Tang et al~\cite{Tang2022}. It enables continuous training on edge \acp{fpga} using a unified channel-level parallelism-based convolutional kernel and a data-reshaping approach. 
Another work targeting edge devices was published by Hong et al. in 2021~\cite{Hong2021}. They designed a lightweight and power-efficient training accelerator for \acp{cnn}. With a focus on low-power for mobile and edge computing, an energy reduction of over \num{4.5} times was reported as compared to existing \ac{fpga} accelerators for the MNIST dataset. 

\Ac{fpga} training accelerators targeting communication systems were presented by Ney et al. in 2022~\cite{Ney2022} and 2023~\cite{Ney2023}. In~\cite{Ney2022}, a trainable \ac{nn}-based demapper was implemented using a cross-layer-design methodology, and in~\cite{Ney2023} a novel unsupervised equalizer was presented featuring a fully-pipelined hardware architecture to balance the lifetime of feature maps. However, none of the works is able to satisfy high throughput requirements surpassing \SI{10}{\giga \bit \per \second}. 

In summary, some works focus on accelerating the throughput and processing time of \ac{nn} training on \ac{fpga}~\cite{Venkataramanaiah2019, Liu2017, Mazouz2021}. Other works address low-power and low-energy processing for accelerating the training at the edge~\cite{Tang2022, Hong2021, Ney2022, Ney2023}. However, there is a clear lack of research focusing on high-throughput inference combined with resource-efficient training on \ac{fpga} which is mandatory for our application to satisfy the high-throughput constraints while providing the required adaptability.

\subsection{\ac{nn}-based Equalization on \ac{fpga}}

Recent research strongly focuses on enhancing the performance of communication systems using \ac{nn}-based algorithms~\cite{zerguine2001, schaedler2019, Ney2022}. However, only a few works analyze the implementation complexity on platforms like \acp{fpga} or \acp{asic} which is an essential step for the deployment of practical communication systems. 

In \cite{Freire2022}, the \ac{fpga} implementations of \ac{rnn} and \ac{cnn}-based equalizers were presented. As a result, the combination of \ac{rnn} and \ac{cnn} provided similar performance as the \ac{dbp} equalizer while achieving a gain compared with the chromatic dispersion compensation baseline. Moreover, the hardware complexity for achieving a \SI{200}{\giga bit \per \second} and a \SI{400}{\giga bit \per \second} throughput was evaluated. 
Another high-throughput \ac{fpga} implementation of a \ac{cnn}-based equalizer was proposed in 2023~\cite{Ney2023_2}. The implementation features optimization from the algorithm down to the hardware architecture. This way, the \ac{cnn}-based equalizer achieved a \ac{ber} one order of magnitude lower than a conventional one while satisfying the throughput requirements of \SI{40}{\giga Bd}. 
Kaneda et al. proposed an \ac{fpga} implementation of an \ac{nn}-based equalizer for high-speed \acp{pon} and showed that the \ac{nn}-based equalizer outperforms the \ac{mlse} equalizer. On \ac{fpga}, the impact of fixed-point resolution was analyzed and it was shown that reducing the weight width from \num{8} to \num{4} bit resulted in a reduction of \SI{40}{\percent} in \ac{lut} resources.   
In~\cite{Li2021} an \ac{fpga} implementation of a parallel pruned \ac{nn} equalizer was presented. A significant reduction in \ac{ber} was demonstrated for a \SI{100}{\giga bit \per \second} channel.  Further, the novel pruning strategy resulted in a \SI{50}{\percent} resource reduction without a degradation of communication performance.  
Huang et al. compared an \ac{rnn} \ac{fpga} implementation with parallel outputs to a fully connected \ac{nn} for equalization of an \ac{imdd} channel~\cite{Huang2022}. They showed that their novel \ac{rnn} structure can outperform the fully connected \ac{nn} with the same input size and the same number of training parameters.

To conclude, several works focus on high-throughput \ac{fpga} implementations of \ac{nn}-based equalizers. However, all of the works only include the inference of the \ac{nn} on the \ac{fpga}. Yet, for changing environmental conditions or varying channel characteristics fast adaption of the equalizer is of high importance. Our work focuses on this gap, where adaptation is required while satisfying the high-throughput constraints. 

\section{System Model}
\label{sec:system_model}

\begin{figure}[t]
	\centerline{\tikzsetnextfilename{SystemModel}
\begin{tikzpicture}[node distance=0.2,>=latex, scale=0.9]
    \pgfdeclarelayer{background}
    \pgfdeclarelayer{foreground}
    \pgfsetlayers{main,foreground}

    \def\minWid{0.6cm}; \def\minHei{0.5cm}; \def\arrowLen{0.5cm};
    \def\boxFontSize{\footnotesize}
    
    \begin{pgfonlayer}{foreground}
    \node[overlay] (in) {};
    \node[block, rounded corners, minimum width=\minWid, minimum height=\minHei] (transmitter) {T};

	\node[block, rounded corners, right=\arrowLen of transmitter, minimum width=\minWid] (txfilter) {$\bm{h}_\text{ps}$};
	\node[block, rounded corners, right=\arrowLen of txfilter, minimum width=\minWid] (channel) {$\bm{h}_\text{ch}$};
	\node[block, rounded corners, right=\arrowLen of channel, minimum width=\minWid] (sld) {$f_{\mathrm{rx}}\mlr{\cdot}$};
	\node[draw, circle,inner sep=-0.0pt, right=\arrowLen of sld] (noise) {$\mathbf{+}$};

    \node[block, rounded corners, minimum width=\minWid, minimum height=\minHei, right=\arrowLen+0.2cm of noise] (equalizer) {EQ};
    \node[block, rounded corners, minimum width=\minWid, minimum height=\minHei, right=\arrowLen of equalizer] (decision) {D};

	\draw[-{Latex[length=2mm]}, thick] (transmitter) -- node[midway, above, xshift=-0.1cm] {$\bm{x}$} (txfilter);
	\draw[-{Latex[length=2mm]}, thick] (txfilter) -- (channel);
	\draw[-{Latex[length=2mm]}, thick] (channel) -- node[midway, above] {$\bm{\tilde{x}}$} (sld);
	\draw[-{Latex[length=2mm]}, thick] (sld) -- (noise);
	\draw[{Latex[length=2mm]}-, thick] (noise) -- node[midway, right, yshift=+0.2cm] {$\bm{n}$} +(0,1.5*\arrowLen);
	\draw[-{Latex[length=2mm]}, thick] (noise.east) -- node[midway, above] {$\bm{y}$}  (equalizer.west);
	\draw[-{Latex[length=2mm]}, thick] (equalizer) -- node[midway, above] {$\bm{z}$} (decision);
    \draw[-{Latex[length=2mm]}, thick] (decision.east) -- node[midway, above, xshift=+0.2cm] {$\bm{\hat{x}}$} +(\arrowLen,0);
    \end{pgfonlayer}

    \node[draw, RPTU_Violett, thick, dotted, rounded corners, inner xsep=0.1cm, inner ysep=0.15cm, fit=(txfilter) (channel) (sld) (noise)] (total_channel) {};
    \node[fill=white] at (total_channel.south) {\textcolor{RPTU_Violett}{\footnotesize Channel}};

    \node[draw, RPTU_GreenGray, thick, dotted, rounded corners, inner xsep=0.1cm, inner ysep=0.15cm, fit=(equalizer) (decision)] (receiver) {};
    \node[fill=white] at (receiver.south) {\textcolor{RPTU_GreenGray}{\footnotesize Receiver}};

\end{tikzpicture}}
	\caption{Model of the communication chain: Data symbols $x$ are sent from a transmitter \textit{T} over a channel to a receiver \textit{R}.}
	\label{fig:sim_channel}
\end{figure}
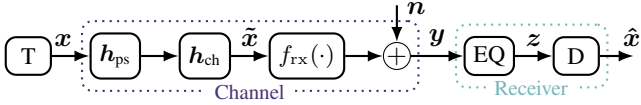

An overview of the simulated system model of this work is given in Fig. \ref{fig:sim_channel}. It consists of the transmitter T, the channel model, and the receiver. The symbols $\bm{x}$ are transmitted over the channel including the pulse shaping filter $\bm{h}_\mathrm{ps}$, the channel impulse response $\bm{h}_\mathrm{ch}$, and the receiver characteristics $f_\mathrm{rx}\mlr{\cdot}$. At the receiver side, the \ac{nn}-based equalizer \textit{EQ} compensates for the distortions introduced by the channel, and a decision \textit{D} is taken based on the equalizer's output.

\subsection{Communication Scenario}
\label{sec:channel_model}
As a potential use-case of our high-speed \ac{nn}-based equalizer we consider \ac{imdd} \acp{pon}. They are usually composed of a point-to-multipoint optical-fiber network such that a single \ac{olt} is connected to multiple endpoints referred to as \acp{onu}. In such systems, increasing data rates lead to severe \ac{isi} and nonlinear distortions~\cite{Simon2020}. To compensate for those distortions equalizers are deployed at the receiver side. Since there is a huge variation in channel conditions for the \acp{onu}~\cite{Elwan2021}, it is not practical to provide a unified equalizer with common parameters for each \ac{onu}. Instead, each equalizer should be adaptable to the specific channel characteristics, which might also fluctuate over time. For this reason, this use-case fits well to our adaptable \ac{nn}-based equalization approach. 

In particular, we consider an optical channel with \ac{imdd} and \ac{pam} as described in \cite{plabst_wiener}. The square-law detection is modeled by $\tilde{\bm{y}} = f_\mathrm{rx}\mlr{\tilde{\bm{x}}}$ with $\tilde{y}_i=|\tilde{x}_i|^2$ and distorts the signal nonlinearly while linear distortions are caused by \ac{cd} given by the frequency response 
\begin{equation*}
    H_\mathrm{cd}\left( L_{\mathrm{fiber}}, f \right) = \exp\left(- \frac{1}{2} \alpha L_{\mathrm{fiber}} \, + \, \mathrm{j} 2 \pi^2 \beta_2 f^2 L_{\mathrm{fiber}}\right)
\end{equation*}
where $L_{\mathrm{fiber}}$ is the fiber length, $\beta_2=-\frac{\lambda^2}{2\pi \mathrm{c}} D_\mathrm{cd}$ is defined by the wavelength $\lambda$, the speed of light $\mathrm{c}$ and the fiber's dispersion coefficient $D_\mathrm{cd}$; $\alpha$ is the fiber attenuation.

For our experiments, we consider C-band transmission at $\lambda=\SI{1550}{\nano\meter}$ with a length of \SI{35}{\kilo \meter}, a \ac{snr} of \SI{15}{\decibel}, $D_\mathrm{cd} = \SI{17}{\CD}$ and $\alpha \triangleq \SI{0.2}{\FL}$. Further, we set the transmission rate to \SI{20}{\giga Bd} since fiber nonlinearities only become significant for high data rates, making nonlinear equalizers like \acp{nn} only relevant for high-data-rate scenarios. Thus, achieving the throughput of \SI{20}{\giga Bd} is a mandatory requirement for our hardware implementation to process the received data in real time.

\subsection{NN Topology}
Our \ac{nn} model of the equalizer employs a 1D \ac{cnn} architecture, based on the analysis conducted in \cite{Ney2023_2}. Specifically, it consists of two \ac{conv1d} layers with a kernel size of \num{9} and a padding of \num{4} where the first layer is followed by a \ac{relu} activation function. The first layer has \num{1} input channel, \num{4} output channels, and a stride of \num{8}, while the second layer has \num{4} input channels, \num{8} output channels and a stride of \num{2}, corresponding to the oversampling factor. This way, \num{8} samples can be calculated jointly in a single pass of the \ac{nn}. 
The \ac{nn} is designed in a detailed trade-off analysis, similar to the one presented in \cite{Ney2023_2}, where further reduction in complexity significantly increases the \ac{ber} of the system.

\section{Hardware Architecture}
\label{sec:hardware_architecture}

In the following, the \ac{fpga} hardware architecture of our \ac{cnn}-based equalizer is described. In contrast to many previous works, the implementation does not only include the inference of the \ac{cnn} but also the training is implemented. This enables retraining of the \ac{cnn}-based equalizer for unseen channels or changing channel conditions. The main focus of the hardware architecture lies on satisfying the high throughput requirements of the optical communication channel which is especially challenging due to the high computational complexity of the backward pass. To reduce the complexity of the hardware implementation, weights, bias, activations, and gradients are represented as arbitrary-width fixed-point numbers. The exact integer and fraction bit widths are determined in a detailed quantization exploration such that neither the training behavior nor the communication performance is sacrificed significantly. 

\subsection{Convolutional Layer}
\label{sec:convolutional_layer}

\begin{figure}[t]
	\centerline{\tikzsetnextfilename{weight_reuse}
\begin{tikzpicture}[node distance=0.2,>=latex, font=\scriptsize]
    \tikzset{near start abs/.style={xshift=.01cm}}
    \def\minWid{3cm}; \def\minHei{0.5cm}; \def\arrowLen{0.4cm};
    \def\boxFontSize{\scriptsize}
    \def\lightColour{50}

    \node[block, draw=black!60, text=black!60, align=left, fill=white, font=\footnotesize, rounded corners=3pt] (C_no_reuse_i4) 
    {
     \hspace{1.3cm} \textbf{Default Conv1D}\\
     \textbf{for} s \textbf{in} $S$:\\
    \hspace{0.3cm} $\forall o \in [\![0,O_\mathrm{c}]\!], i \in [\![0,I_\mathrm{c}]\!]$\\[0.1cm]
    \hspace{0.3cm} $\mathbf{Y}_\textrm{0, o, s} = \sum_{k=-\frac{K}{2}}^{\frac{K}{2}} \mathbf{X}_\textrm{0, i, s+k} \cdot \mathbf{W}_\textrm{i, o, k}$
    };

    \node[block, draw=black, align=left, minimum width=1cm,  fill=white, below=-1.5cm of C_no_reuse_i4, xshift=0.4cm, font=\footnotesize, rounded corners=3pt] (C_no_reuse_i1) 
    {
    \hspace{1.3cm} \textbf{Default Conv1D}\\
     \textbf{for} s \textbf{in} $S$:\\
    \hspace{0.3cm} $\forall o \in [\![0,O_\mathrm{c}]\!], i \in [\![0,I_\mathrm{c}]\!]$\\[0.1cm]
    \hspace{0.3cm} $\mathbf{Y}_\textrm{0, o, s} = \sum_{k=-\frac{K}{2}}^{\frac{K}{2}} \mathbf{X}_\textrm{0, i, s+k} \cdot \mathbf{W}_\textrm{i, o, k}$
    };
    
    \path ($(C_no_reuse_i4.south west)+(0,+0.03cm)$) edge [draw=none] node [sloped, anchor=center] {\normalsize ...} ($(C_no_reuse_i1.south west)+(0,+0.03cm)$);
    \path ($(C_no_reuse_i4.north east)+(-0.04,-0.03cm)$) edge [draw=none] node [sloped, anchor=center] {\normalsize ...} ($(C_no_reuse_i1.north east)+(-0.04,-0.03cm)$);
    \path ($(C_no_reuse_i4.north west)+(0.02,-0.04cm)$) edge [draw=none] node [sloped, anchor=center] {\normalsize ...} ($(C_no_reuse_i1.north west)+(0.02,-0.04cm)$);

    \node[left=1.5*\arrowLen of C_no_reuse_i4, yshift=+0.45cm, text=black!\lightColour, font=\footnotesize] (X4) {\small $\mathbf{X}_{\mathrm{B}-1}$};
    \node[left=1.5*\arrowLen of C_no_reuse_i1, yshift=+0.15cm, font=\footnotesize] (X1) {\small $\mathbf{X}_0$};

    \node[left=1.5*\arrowLen of C_no_reuse_i4, yshift=+0.15cm, text=black!\lightColour, font=\footnotesize] (W4) {\small \textbf{W}};
    \node[left=1.5*\arrowLen of C_no_reuse_i1, yshift=-0.15cm, font=\footnotesize] (W1) {\small \textbf{W}};

    \node[right=1.5*\arrowLen of C_no_reuse_i4, text=black!\lightColour, font=\footnotesize] (Y4) {\small $\mathbf{Y}_{\mathrm{B}-1}$};
    \node[right=1.5*\arrowLen of C_no_reuse_i1, font=\footnotesize] (Y1) {\small $\mathbf{Y}_0$};

    \draw[->] (X1) -- ($(C_no_reuse_i1.west)+(0,0.15cm)$);
    \draw[->, black!\lightColour] (X4) -- ($(C_no_reuse_i4.west)+(0,0.45cm)$);

    \draw[->] (W1) -- ($(C_no_reuse_i1.west)+(0,-0.15cm)$);
    \draw[->, black!\lightColour] (W4) -- ($(C_no_reuse_i4.west)+(0,+0.15cm)$);
    
    \begin{pgfonlayer}{background}
        \draw[->] (C_no_reuse_i1) -- (Y1);
        \draw[->, black!\lightColour] (C_no_reuse_i4) -- (Y4);
    \end{pgfonlayer}

    \node[block, draw=black, align=left, minimum width=5cm, fill=white, below=0.5cm of C_no_reuse_i1, font=\footnotesize, rounded corners=3pt] (C_multi_inst) 
    {
     \hspace{1.3cm}\textbf{Batch-Parallel Conv1D}\\
     \textbf{for} s \textbf{in} $S$:\\
     \hspace{0.5cm} $\forall n \in [\![0,B]\!], o \in [\![0,O_\mathrm{c}]\!], i \in [\![0,I_\mathrm{c}]\!]$\\[0.1cm]
    \hspace{0.5cm} $\mathbf{Y}_\textrm{n, o, s} = \sum_{k=-\frac{K}{2}}^{\frac{K}{2}} \mathbf{X}_\textrm{n, i, s+k} \cdot \mathbf{W}_\textrm{i, o, k}$

    };

    \node[left=1*\arrowLen of C_multi_inst, yshift=+0.6cm, font=\footnotesize] (X1_2) {\small $\mathbf{X}_0$};
    \node[left=1*\arrowLen of C_multi_inst, yshift=+0.2cm] (vdots_in_2) {\small $\vdots$};
    \node[left=1*\arrowLen of C_multi_inst, yshift=-0.3cm, font=\footnotesize] (X4_2) {\small $\mathbf{X}_{\mathrm{B}-1}$};
    \node[left=1*\arrowLen of C_multi_inst, yshift=-0.6cm, font=\footnotesize] (W_2) {\small \textbf{W}};

    \node[right=1*\arrowLen of C_multi_inst, yshift=+0.6cm, font=\footnotesize] (Y1_2) {\small $\mathbf{Y}_0$};
    \node[right=1*\arrowLen of C_multi_inst, yshift=+0.15cm] (vdots_out_2) {\small $\vdots$};
    \node[right=1*\arrowLen of C_multi_inst, yshift=-0.6cm, font=\footnotesize] (Y4_2) {\small $\mathbf{Y}_{\mathrm{B}-1}$};
    
    \draw[->] (X1_2) -- ($(C_multi_inst.west)+(0,0.6cm)$);
    \draw[->] (X4_2) -- ($(C_multi_inst.west)+(0,-0.3cm)$);
    \draw[->] (W_2) -- ($(C_multi_inst.west)+(0,-0.6cm)$);

    \begin{pgfonlayer}{background}
        \draw[->] ($(C_multi_inst.east)+(0,0.6cm)$) -- (Y1_2);
        \draw[->] ($(C_multi_inst.east)+(0,-0.6cm)$) -- (Y4_2);
    \end{pgfonlayer}
    
\end{tikzpicture}}
	\caption{Comparison of default \ac{conv1d} and batch-parallel \ac{conv1d} architecture. All operations in the loops are processed in parallel in one clock cycle.}
 	\label{fig:conv_weight_resuse}
\end{figure}
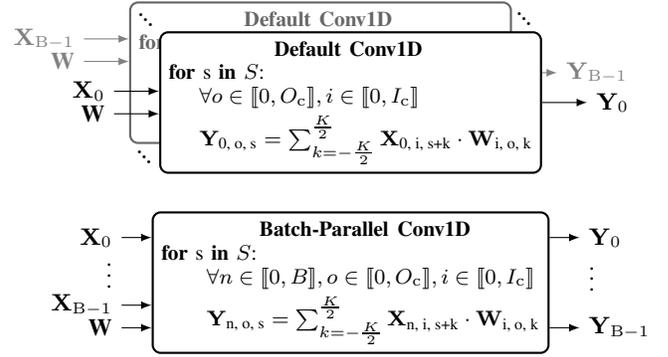

Our hardware architecture is based on fully parallelized \ac{conv1d} layers to maximize the throughput of the implementation. Mathematically, the \ac{conv1d} layer is expressed as: 
\begin{align}
\begin{split}
        \mathbf{Y}_\textrm{n, o, s} = &\sum_{i=0}^{I_\mathrm{c}} \sum_{k=-\frac{K}{2}}^{\frac{K}{2}} \mathbf{X}_\textrm{n, i, s+k} \cdot \mathbf{W}_\textrm{i, o, k}\\ 
        &\forall n \in [\![0,B]\!], o \in [\![0,O_\mathrm{c}]\!], s \in [\![0,S]\!]
    \end{split}
    \label{eq:conv1d}
\end{align}
where $\mathbf{X}$ is the input tensor with shape $N \times I_\mathrm{c} \times S$, $\mathbf{Y}$ is the output tensor with shape $N \times O_\mathrm{c} \times S$, $\mathbf{W}$ is the weight with shape $I_\mathrm{c} \times  O_\mathrm{c} \times K$, $I_\mathrm{c}$ gives the number of input channels, $O_\mathrm{c}$ gives the number of output channels, $K$ is the kernel size, $S$ is the sequence length, $B$ is the batch size, and $[\![0,N]\!]$ gives the set of natural numbers from $0$ to $N$. 

The architecture of our \ac{conv1d} layer is based on the one presented in \cite{Ney2023}, which is parallelized on the level of input channels, output channels, and on the kernel level. However, in contrast to \cite{Ney2023}, where one \ac{cnn} module is instantiated for each sample in a batch of size $B$, we include the batch-level parallelism inside the \ac{conv1d} layer, as shown in Fig. \ref{fig:conv_weight_resuse}. 
Thus, the weights can be reused for each of the $B$ samples in one batch. This way, the inputs of each instance are first multiplied and accumulated with the same weight before moving the convolutional kernel forward. Eventually, this reduces the resource usage of our architecture, as shown in Sec. \ref{sec:hardware_architecture_analysis}. Further, including the batch-parallelism in the \ac{conv1d} layer enables our custom \ac{dsp}-mapping scheme, as presented in Sec. \ref{sec:dsp_mapping}.

\subsection{DSP Mapping}
\label{sec:dsp_mapping}

\begin{figure}[t]	\centerline{\tikzsetnextfilename{dsp_mapping_operands}
\begin{tikzpicture}[font=\scriptsize]
    \def\minWid{1.1cm}; \def\minHei{0.5cm}; \def\arrowLen{1cm};

    \node[block, draw=black, minimum width=\minWid, minimum height=0.4cm, fill=white, inner sep=0cm] (D1) {$D_\mathrm{1}$};
    \node[block, draw=black, minimum width=\minWid, , minimum height=0.4cm, fill=white, inner sep=0cm,  left=0cm of D1, xshift=+0.25mm] (w_zeros) {00...00};
    \node[block, draw=black, minimum width=\minWid, , minimum height=0.4cm, fill=white, inner sep=0cm,  left=0cm of w_zeros, xshift=+0.25mm] (D2) {$D_\mathrm{2}$};
    \node[block, draw=black, minimum width=0.5cm, , minimum height=0.4cm, fill=white, inner sep=0cm,  left=0cm of D2, xshift=+0.25mm] (zero) {0};

    \node[below=0.1cm of w_zeros, xshift=-0.3cm] (D) {\small \textbf{D}};
    
    \draw [|-|] ($(D1.west) + (0, 0.4cm)$) -- node [above] {$d$ bit} ($(D1.east) + (0, 0.4cm)$);

    \draw [|-|] ($(w_zeros.west) + (0, 0.4cm)$) -- node [above] {$w$ bit} ($(w_zeros.east) + (0, 0.4cm)$);

    \draw [|-|] ($(D2.west) + (0, 0.4cm)$) -- node [above] {$d$ bit} ($(D2.east) + (0, 0.4cm)$);

    \draw [|-|] ($(zero.west) + (0, 0.4cm)$) -- node [above] {$1$ bit} ($(zero.east) + (0, 0.4cm)$);

    \node[right=0.05cm of D1] (X) {\Large $\mathbf{\cdot}$};

    \node[block, draw=black, minimum width=\minWid, , minimum height=0.4cm, fill=white, inner sep=0cm,  right=0.05cm of X] (W) {W};
    
    \node[below=0.1cm of W] (W_name) {\small \textbf{W}};

    \draw [|-|] ($(W.west) + (0, 0.4cm)$) -- node [above] {$w$ bit} ($(W.east) + (0, 0.4cm)$);

    \node[left=0.05cm of zero] (equal) {\large $\mathbf{=}$};
    
    \node[block, draw=black, minimum width=1.2cm, minimum height=0.4cm, fill=white, inner sep=0cm, left=0.05cm of equal] (R1) {$R_\mathrm{1}$};
    \node[block, draw=black, minimum width=1.2cm, , minimum height=0.4cm, fill=white, inner sep=0cm,  left=0cm of R1, xshift=+0.25mm] (R2) {$R_\mathrm{2}$};

    \draw [|-|] ($(R1.west) + (0, 0.4cm)$) -- node [above] {$d + w$ bit} ($(R1.east) + (0, 0.4cm)$);
    \draw [|-|] ($(R2.west) + (0, 0.4cm)$) -- node [above] {$d + w$ bit} ($(R2.east) + (0, 0.4cm)$);

    \node[below=0.1cm of R1.west, yshift=-0.2cm] (R_name) {\small \textbf{R}};

\end{tikzpicture}}
	\caption{Mapping of the two multiplication $R_\mathrm{1} = D_\mathrm{1} \cdot W$ and $R_\mathrm{2} = D_\mathrm{2} \cdot W$ to a single \ac{dsp}}
 	\label{fig:dsp_mapping_operands}
\end{figure}
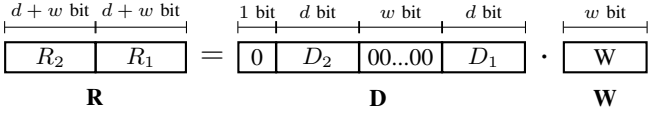

To efficiently utilize the constrained \ac{dsp} resources of the \ac{fpga}, we present a novel approach to map two signed\mbox{-}unsigned multiplications of the second \ac{conv1d} layer of our \ac{cnn} to one \ac{dsp}. As explained in Sec. \ref{sec:convolutional_layer}, our \ac{conv1d} layer architecture processes multiple samples of one batch of size $B$ in an unrolled conv1d block. Thus, different inputs are multiplied by the same weight. This allows to reuse each weight $B$ times. In the following, we will refer to the input data $X_{n, i, s+k}$  of equation (\ref{eq:conv1d}) as $D_\mathrm{1}$ and to $X_{n+1, i, s+k}$ as $D_\mathrm{2}$ where the width of the datatype is given as $d$. $W$ is short for the weight $W_{i, o, k}$ with width $w$, and the results of the multiplications are given as $R_\mathrm{1}$ and $R_\mathrm{2}$ with width $r$.  Our goal is to map the two multiplications $D_\mathrm{1} \cdot W = R_\mathrm{1}$ and $D_\mathrm{2} \cdot W = R_\mathrm{2}$ to a single \ac{dsp}. 



\subsubsection{Constraints}
\label{sec:constraints}

Even with the similar operand $W$, further constraints must be applied to map the operations efficiently to one \ac{dsp}. One important constraint is the signedness of the operands. 
In this case of two unsigned operands, inserting $w$ guard bits between the operands $D_\mathrm{1}$ and $D_\mathrm{2}$ is sufficient. Then the first $d+w=r$ bits of the result $R$ correspond to $R_\mathrm{1}$ and the second $r$ bits correspond to $R_\mathrm{2}$. However, for our application, $W$ is of type signed since it can also take negative values while $D_\mathrm{1}$ and $D_\mathrm{2}$ are of type unsigned for the second \ac{conv1d} layer since it's preceded by a \ac{relu} activation function. This problem is slightly different from the one considered in \cite{Nguyen2017, Lee2019}, where the common operand is of type unsigned. Therefore, the solution of \cite{Nguyen2017, Lee2019} isn't applicable in our case. Another important constraint is the bit width of the operands. The \textit{DSP48-E2} block of our target \ac{fpga} features a $27 \times 18$ bit multiplier, thus our combined data input $D$ is limited to \SI{27}{\bit}. As shown in Fig. \ref{fig:dsp_mapping_operands}, $D$ consists of $D_\mathrm{1}$ and $D_\mathrm{2}$, $w$ guard bits, and one leading zero. Thus, $2\cdot w + d \leq 26$. Experiments showed, that under this constraint a configuration with \SI{6}{\bit} for the weights and \SI{10}{\bit} for the inputs provides the best performance. 

In summary, to enable our approach of mapping two multiplications to one \ac{dsp} the following constraints need to be satisfied:
\begin{itemize}
    \item A common multiplicand $W$ is used in both multiplication
    \item $W$ is of type signed, while the other operands $D_\mathrm{1}$ and $D_\mathrm{2}$ are of type unsigned. 
    \item $2\cdot w + d \leq 26$
\end{itemize}

\subsubsection{Modification}
For the combined multiplication, the result of $D_\mathrm{1} \cdot W$ potentially influences the result $R_\mathrm{2}$ of the multiplication $D_\mathrm{2} \cdot W$. In the case of $W \geq 0$, $R_\mathrm{1}$ is signed-extended with zeros since $\mathrm{Sign}(W) = 0$ and $\mathrm{Sign}(D_\mathrm{1}) = 0$. Thus, no overflow is propagated into the calculation of $R_\mathrm{2}$. However, when $W < 0$, $\mathrm{Sign}(W) = 1$ and $\mathrm{Sign}(R_\mathrm{1}) = 1$. Thus, $R_\mathrm{1}$ is signed-extended with ones, which are added to the calculation of $R_\mathrm{2}$. If we declare $\Delta R_\mathrm{2}$ as the difference between the correct result $R_\mathrm{2}$ and the \ac{dsp} output $R_\mathrm{2}^*$ then:
\begin{equation*}
    \Delta R_\mathrm{2} = R_\mathrm{2} - R_\mathrm{2}^* = 111...111_\mathrm{2} = -1_{\mathrm{10}} \;   .
\end{equation*}
Thus, to get the correct result $R_\mathrm{2}$, we need to add $1$ to $R_\mathrm{2}^*$.

To summarize, the following steps are required to map two signed-unsigned multiplications to a single \ac{dsp}, as implemented in our hardware architecture:
\begin{enumerate}
    \item get $D$ by concatenating $D_\mathrm{1}$ and $D_\mathrm{2}$ with $w$ guard bits
    \item extend $D$ with zeros and sign-extend $W$
    \item multiply $D$ with $W$ using a single \ac{dsp}
    \item the first $d+w$ bits corresponds to the result $R_\mathrm{1}$
    \item the second $d+w$ bits correspond to $R_\mathrm{2}^*$
    \item if $W < 0$ then $R_\mathrm{2} = R_\mathrm{2}^* + 1$ else $R_\mathrm{2} = R_\mathrm{2}^*$
\end{enumerate}

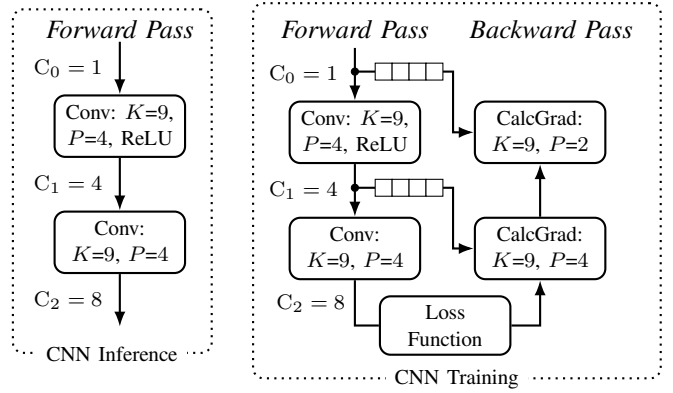
\begin{figure}[t]	\centerline{\tikzsetnextfilename{InferenceAndTrainingModule}
\begin{tikzpicture}[node distance=0.2,>=latex, scale=0.9]
    \tikzset{near start abs/.style={xshift=.01cm}};
    \def\minWid{3.4cm}; \def\minHei{0.5cm}; \def\arrowLen{0.7cm};
    \def\boxFontSize{\footnotesize}
    
    \def\colLight{15};
    \def\boxCol{RPTU_BlueGray}; \def\DotBoxCol{RPTU_LightGreen};

    \node[text=black, inner sep=0.1em] (forward_text_inference) {\textit{Forward Pass}};

    \node[block, draw=black, rounded corners, minimum height=\minHei, below=\arrowLen*1.0 of forward_text_inference, text width=1.5cm] (conv0_inference) {\boxFontSize Conv: $K$=$9$, $P$=$4$, ReLU};

    \node[block, draw=black, rounded corners,  minimum height=\minHei, below=\arrowLen of conv0_inference, text width=1.5cm] (conv1_inference) {\boxFontSize Conv: $K$=$9$, $P$=$4$};

    \node[below=\arrowLen of conv1_inference] (output_inference) {};

    \draw[-{Latex[length=2mm]}, thick] (forward_text_inference) -- node[midway, left, xshift=-0.1cm] {\footnotesize $\mathrm{C}_0=1$} (conv0_inference.north);
    \draw[-{Latex[length=2mm]}, thick] (conv0_inference) -- node[midway, left, xshift=-0.1cm] {\footnotesize $\mathrm{C}_1=4$} (conv1_inference.north);
    \draw[-{Latex[length=2mm]}, thick] (conv1_inference.south) -- node[midway, left, xshift=-0.1cm] {\footnotesize $\mathrm{C}_2=8$} (output_inference);

    \node[draw, black, thick, dotted, rounded corners, inner xsep=0.3cm, inner ysep=0.1cm, xshift=-0.1cm, fit=(forward_text_inference) (conv0_inference) (conv1_inference) (output_inference)] (inference) {};
    \node[fill=white] at (inference.south) {\textcolor{black}{\footnotesize CNN Inference}};

    \node[text=black, right=1cm of forward_text_inference] (forward_text) {\textit{Forward Pass}};
    \node[right=0.3cm of forward_text, text=black] (backward_text) {\textit{Backward Pass}};


    \node[block, draw=black, rounded corners,  minimum height=\minHei, below=\arrowLen*1.0 of forward_text, text width=1.5cm] (conv0) {\boxFontSize Conv: $K$=$9$, $P$=$4$, ReLU};

    \node[block, draw=black, rounded corners, minimum height=\minHei, below=\arrowLen of conv0, text width=1.5cm] (conv1) {\boxFontSize Conv: $K$=$9$, $P$=$4$};

    \node[block, draw=black, rounded corners, minimum height=\minHei, below=0.2cm of conv1, xshift=1.2cm, text width=1.5cm] (sup_loss) {\boxFontSize Loss Function};



    
    \node[block, draw=black, rounded corners, right=0.7cm of conv1, text width=1.5cm] (grad1) {\boxFontSize CalcGrad: $K$=$9$, $P$=$4$};

    \node[block, draw=black, rounded corners,  right=0.7cm of conv0, yshift=0.0cm, text width=1.5cm] (grad0) {\boxFontSize CalcGrad: $K$=$9$, $P$=$2$};

    \draw[-{Latex[length=2mm]}, thick] (forward_text) -- node[midway, left, xshift=-0.1cm] {\footnotesize $\mathrm{C}_0=1$} (conv0.north);
    \filldraw (forward_text.south) +(0,-\arrowLen*0.5) circle (1.5pt);

    \draw[-{Latex[length=2mm]}, thick] (conv0) -- node[midway, left, xshift=-0.1cm] {\footnotesize $\mathrm{C}_1=4$} (conv1.north);
    \filldraw (conv0.south) +(0,-\arrowLen*0.5) circle (1.5pt);

    \draw[-{Latex[length=2mm]}, thick] (forward_text.south)++(0, -\arrowLen*0.5) -- +(1.45cm, 0) |- (grad0.west);
    \draw[-{Latex[length=2mm]}, thick] (conv0.south)++(0, -\arrowLen*0.5) -- +(1.45cm, 0) |- (grad1.west);

    \def\recSize{0.25cm};
    
    \filldraw[draw=black, fill=white] (forward_text.south)++(0.3cm, -\arrowLen*0.5 + \recSize * 0.5) rectangle ++(\recSize,-\recSize);
    \filldraw[draw=black, fill=white] (forward_text.south)++(0.3cm + 1 * \recSize, -\arrowLen*0.5 + \recSize * 0.5) rectangle ++(\recSize,-\recSize);
    \filldraw[draw=black, fill=white] (forward_text.south)++(0.3cm + 2 * \recSize, -\arrowLen*0.5 + \recSize * 0.5) rectangle ++(\recSize,-\recSize);
    \filldraw[draw=black, fill=white] (forward_text.south)++(0.3cm + 3 * \recSize, -\arrowLen*0.5 + \recSize * 0.5) rectangle ++(\recSize,-\recSize);

    \filldraw[draw=black, fill=white] (conv0.south)++(0.3cm, -\arrowLen*0.5 + \recSize * 0.5) rectangle ++(\recSize,-\recSize);
    \filldraw[draw=black, fill=white] (conv0.south)++(0.3cm + 1 * \recSize, -\arrowLen*0.5 + \recSize * 0.5) rectangle ++(\recSize,-\recSize);
    \filldraw[draw=black, fill=white] (conv0.south)++(0.3cm + 2 * \recSize, -\arrowLen*0.5 + \recSize * 0.5) rectangle ++(\recSize,-\recSize);
    \filldraw[draw=black, fill=white] (conv0.south)++(0.3cm + 3 * \recSize, -\arrowLen*0.5 + \recSize * 0.5) rectangle ++(\recSize,-\recSize);

    \draw[-, thick] (conv1) |- node[near start, left] {\footnotesize $\mathrm{C}_2=8$}  (sup_loss.west);
    \draw[-{Latex[length=2mm]}, thick] (sup_loss.east) -| (grad1.south);




    \draw[-{Latex[length=2mm]}, thick] (grad1.north) -- (grad0.south);

    \node[draw, black, thick, dotted, rounded corners, inner xsep=0.25cm, inner ysep=0.2cm, yshift=-0.1cm, fit=(forward_text) (backward_text) (conv0) (conv1) (sup_loss) (grad1) (grad0)] (training) {};
    \node[fill=white] at (training.south) {\textcolor{black}{\footnotesize CNN Training}};
    
\end{tikzpicture}}
	\caption{Comparision of CNN inference and training}
 \label{fig:inference_and_training_module}
\end{figure}

\subsection{Inference and Training Module}
In Fig. \ref{fig:inference_and_training_module} a high-level overview of the \ac{cnn} inference and training module is shown. A similar architecture was presented in \cite{Ney2023}, however in this work, the \ac{cnn} topology is optimized and reduced in size. Further, a separate \ac{cnn} inference module is implemented to satisfy the high throughput requirements. 
The core of the inference module is the highly optimized and parallelized \ac{conv1d} layer. Each layer is connected with an \textit{AXI-Stream} interface to allow for pipelined processing with minimal control overhead. 
For the training module, forward pass and backward pass are also implemented in a pipeline fashion to balance the lifetime of the feature maps such that the memory footprint is reduced, as described in \cite{Ney2023}. Similar to the inference module, the layers of the training module are fully unrolled to enable high-throughput processing. However, batch-parallelism and the custom \ac{dsp} mapping can only be exploited for the inference module, since the constraints of Sec. \ref{sec:constraints} are not satisfied by the training module. 

\subsection{Parallelism of Inference and Training Module}

\begin{figure}[t]
	\centerline{\newcommand{\mydots}{\makebox[0.5em][c]{.\hfil.\hfil.}}

\tikzsetnextfilename{inference_training_partitioning}
\begin{tikzpicture}
    \def\minWid{3cm}; \def\minHei{0.5cm}; \def\arrowLen{1cm};

    \node[block, line width=0.8pt, draw=black, minimum width=1.5cm, minimum height=0.3cm, fill=white, inner sep=0cm] (s_0_0) {};
    \node[block, line width=0.8pt, draw=black, minimum width=1.5cm, , minimum height=0.3cm, fill=white, inner sep=0cm,  right=0cm of s_0_0, xshift=-0.25mm] (s_1_0) {};
    \node[block, line width=0.8pt, draw=black, minimum width=1.5cm, , minimum height=0.3cm, fill=white, inner sep=0cm,  right=0cm of s_1_0, xshift=-0.25mm] (s_2_0) {};
    \node[block, line width=0.8pt, draw=black, minimum width=1.5cm, , minimum height=0.3cm, fill=white, inner sep=0cm,  right=0cm of s_2_0, xshift=-0.25mm] (s_3_0) {};
    \node[block, line width=0.8pt, draw=black, minimum width=1.5cm, , minimum height=0.3cm, fill=white, inner sep=0cm,  right=0cm of s_3_0, xshift=-0.25mm] (s_4_0) {};

    \draw [-, line width=0.8pt] ($(s_0_0.north) - (0, 0.4pt)$) --  ++(-1.0cm, 0);
    \draw [-, line width=0.8pt] ($(s_0_0.south) + (0, 0.4pt)$) -- ++(-1.0cm, 0);

    \draw [-, line width=0.8pt] ($(s_4_0.north) - (0, 0.4pt)$) -- node[at end] (north_east_end) {} ++(1.0cm, 0);
    \draw [-, line width=0.8pt] ($(s_4_0.south) + (0, 0.4pt)$) -- node[at end] (south_east_end) {} ++(1.0cm, 0);

    \node[left=0.1cm of s_0_0] (dots_start) {\large \mydots};
    \node[right=0.1cm of s_4_0] (dots_end) {\large \mydots};

    \node[block, rounded corners, line width=0.8pt, draw=black, minimum width=1.3cm, minimum height=1cm, fill=white, inner sep=0cm,  below=1.0cm of s_0_0] (cnn_train_0) {CNN\\Training};
    \node[block, rounded corners, line width=0.8pt, draw=black, minimum width=1.3cm, minimum height=1cm, fill=white, inner sep=0cm,  below=1.0cm of s_1_0] (cnn_train_1) {CNN\\Training};
    \node[minimum height=1cm, below=1.0cm of s_2_0] (cnn_inference_0_0) {};
    \node[minimum height=1cm, below=1.0cm of s_3_0] (cnn_inference_1_0) {};
    \node[minimum height=1cm, below=1.0cm of s_4_0] (cnn_inference_2_0) {};
    \node[block, rounded corners, line width=0.8pt, draw=black, minimum width=3.9cm, minimum height=1cm, fill=white, inner sep=0cm,  below=1.0cm of s_3_0] (cnn_inference) {CNN\\Inference};

    \node[block, rounded corners, line width=0.8pt, draw=black, fill=white,  below=2.7cm of s_0_0.east] (weights_0) {Weights};

    \draw [-{Latex[length=2mm]}, thick] ($(cnn_train_0.south) + (0.2cm, 0.0cm)$) to [bend right=25] ($(weights_0.north) + (-0.6cm, 0.0cm)$);
    \draw [-{Latex[length=2mm]}, thick]  ($(weights_0.north) + (-0.4cm, 0.0cm)$)  to [bend right=25] ($(cnn_train_0.south) + (0.4cm, 0.0cm)$);

    \draw [-{Latex[length=2mm]}, thick] ($(cnn_train_1.south) + (-0.5cm, 0.0cm)$) to [bend right=25] ($(weights_0.north) + (+0.3cm, 0.0cm)$);
    \draw [-{Latex[length=2mm]}, thick]  ($(weights_0.north) + (0.5cm, 0.0cm)$)  to [bend right=25] ($(cnn_train_1.south) + (-0.3cm, 0.0cm)$);

    \draw [-{Latex[length=2mm]}, thick]  ($(weights_0.east)$)  to [bend right=25] ($(cnn_inference_0_0.south) + (-0.4cm, 0.0cm)$);

    \node[block, line width=0.8pt, draw=black, minimum width=1.5cm, minimum height=0.3cm, fill=white, inner sep=0cm, below=3.6cm of s_0_0] (so_0_0) {};
    \node[block, line width=0.8pt, draw=black, minimum width=1.5cm, , minimum height=0.3cm, fill=white, inner sep=0cm,  right=0cm of so_0_0, xshift=-0.25mm] (so_1_0) {};
    \node[block, line width=0.8pt, draw=black, minimum width=1.5cm, , minimum height=0.3cm, fill=white, inner sep=0cm,  right=0cm of so_1_0, xshift=-0.25mm] (so_2_0) {};
    \node[block, line width=0.8pt, draw=black, minimum width=1.5cm, , minimum height=0.3cm, fill=white, inner sep=0cm,  right=0cm of so_2_0, xshift=-0.25mm] (so_3_0) {};
    \node[block, line width=0.8pt, draw=black, minimum width=1.5cm, , minimum height=0.3cm, fill=white, inner sep=0cm,  right=0cm of so_3_0, xshift=-0.25mm] (so_4_0) {};

    \draw [-, line width=0.8pt] ($(so_0_0.north) - (0, 0.4pt)$) -- ++(-1.0cm, 0);
    \draw [-, line width=0.8pt] ($(so_0_0.south) + (0, 0.4pt)$) -- ++(-1.0cm, 0);

    \draw [-, line width=0.8pt] ($(so_4_0.north) - (0, 0.4pt)$) -- ++(1.0cm, 0);
    \draw [-, line width=0.8pt] ($(so_4_0.south) + (0, 0.4pt)$) -- ++(1.0cm, 0);

    \node[left=0.1cm of so_0_0] (dots_start) {\large \mydots};
    \node[right=0.1cm of so_4_0] (dots_end) {\large \mydots};

    \draw [-{Latex[length=1.5mm]}, densely dashed] (s_0_0) -- (cnn_train_0);
    \draw [-{Latex[length=1.5mm]}, densely dashed] (s_1_0) -- (cnn_train_1);
    \draw [-{Latex[length=1.5mm]}, densely dashed] (s_2_0) -- (cnn_inference_0_0);
    \draw [-{Latex[length=1.5mm]}, densely dashed] (s_3_0) -- (cnn_inference_1_0);
    \draw [-{Latex[length=1.5mm]}, densely dashed] (s_4_0) -- (cnn_inference_2_0);

    \draw [-{Latex[length=1.5mm]}, densely dashed] (cnn_train_0) -- (so_0_0);
    \draw [-{Latex[length=1.5mm]}, densely dashed] (cnn_train_1) -- (so_1_0);
    \draw [-{Latex[length=1.5mm]}, densely dashed] (cnn_inference_0_0) -- (so_2_0);
    \draw [-{Latex[length=1.5mm]}, densely dashed] (cnn_inference_1_0) -- (so_3_0);
    \draw [-{Latex[length=1.5mm]}, densely dashed] (cnn_inference_2_0) -- (so_4_0);

    \draw [|-|, thick] ($(cnn_train_0.west) + (0, 0.8cm)$) -- node[above, inner sep=0.15em](PT) {$\mathrm{P}_\mathrm{T}$} ($(cnn_train_1.east) + (0, 0.8cm)$);
    \draw [|-|, thick] ($(cnn_inference.west) + (0, 0.8cm)$) -- node[above, fill=white, inner sep=0.15em](PI) {$\mathrm{P}_\mathrm{I}$} ($(cnn_inference.east) + (0, 0.8cm)$);

    \node[draw, black, thick, rounded corners, inner xsep=0.15cm, inner ysep=0.1cm, fit=(cnn_train_0) (cnn_inference) (cnn_inference_0_0) (cnn_inference_1_0) (cnn_inference_2_0) (weights_0) (PT) (PI)] (complete_module_0) {};



     


\end{tikzpicture}}
	\caption{Parallelism of the training and the inference modules}
\label{fig:inference_vs_training_partitioning}
\end{figure}
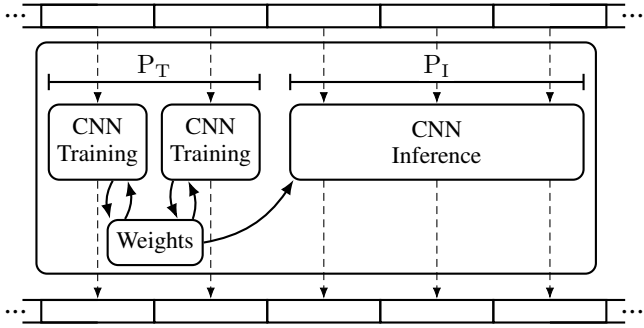

As shown in Fig. \ref{fig:inference_and_training_module}, the \ac{cnn} training is much more complex than the inference since in addition to the forward pass, the loss function, the backward pass, and the weight update need to be implemented. Further, a higher bit width of around \SI{24}{bit} is required as the values of the gradients are usually very small. Therefore, the resource consumption of the \ac{fpga} is much larger for the training module as compared to the inference module with similar parallelism. Thus, in our hardware architecture, we allow for separate \acp{dop} for inference and training, referred to as $\mathrm{P}_\mathrm{I}$ and  $\mathrm{P}_\mathrm{T}$ respectively. In particular, $\mathrm{P}_\mathrm{I}$ is increased by exploiting the batch-level parallelism in the convolutional layer, while $\mathrm{P}_\mathrm{T}$ is increased by using multiple training modules, as shown in Fig. \ref{fig:inference_vs_training_partitioning}. This way, by increasing $\mathrm{P}_\mathrm{I}$ while keeping $\mathrm{P}_\mathrm{T}$ small, we achieve high throughput for the inference, while enabling training of the \ac{cnn}, even on resource-constraint platforms like \acp{fpga}. It's worth mentioning that in the case of a small $\mathrm{P}_\mathrm{T}$, the training is only performed on a subset of the complete input sequence and not all symbols are used for training, however in Sec. \ref{sec:convergence_behavior} we show that stable convergence can be achieved with a low $\mathrm{P}_\mathrm{T}$.

To summarize, our novel hardware architecture includes the following features to satisfy the high throughput requirements: 
\begin{itemize}
    \item The \ac{conv1d} layer is parallelized on the level of input channels, output channels, and on the kernel level.
    \item Batch-level parallelism is exploited by the \ac{conv1d} layer of the inference module to enable weight-reuse.
    \item A custom dsp-mapping scheme is included in the \ac{conv1d} layer to efficiently use the \ac{dsp} resources.
    \item Each layer is implemented as a separate pipeline stage to exploit temporal parallelism.
    \item Different \acp{dop} are used for inference and training module to balance the resources. 
\end{itemize}

\section{Results}

In this section, our \ac{conv1d} hardware architecture is analyzed, the equalizer implementation is evaluated for a high-throughput optical fiber channel, and the \ac{fpga} implementation is compared to state-of-the-art and to an embedded \ac{gpu} implementation.

\subsection{Conv1D Hardware Architecture Analysis}
\label{sec:hardware_architecture_analysis}

In the following, it is analyzed how three different hardware architectures of the \ac{conv1d} layer influence the resource utilization on \ac{fpga}. The first, commonly used, architecture $\mathrm{conv}_\mathrm{def}$ exploits the parallelism of the convolutional layer in terms of input channels, output channels, and on the kernel level. In this case, batch-level parallelism is introduced by instantiating one \ac{cnn} instance for each sample in a batch. The second architecture $\mathrm{conv}_\mathrm{inst}$ includes the batch-parallelism inside the \ac{conv1d} layer, as described in Sec. \ref{sec:convolutional_layer}. The third architecture $\mathrm{conv}_\mathrm{map}$ is the same as $\mathrm{conv}_\mathrm{inst}$, extended by our custom \ac{dsp}-mapping scheme. 

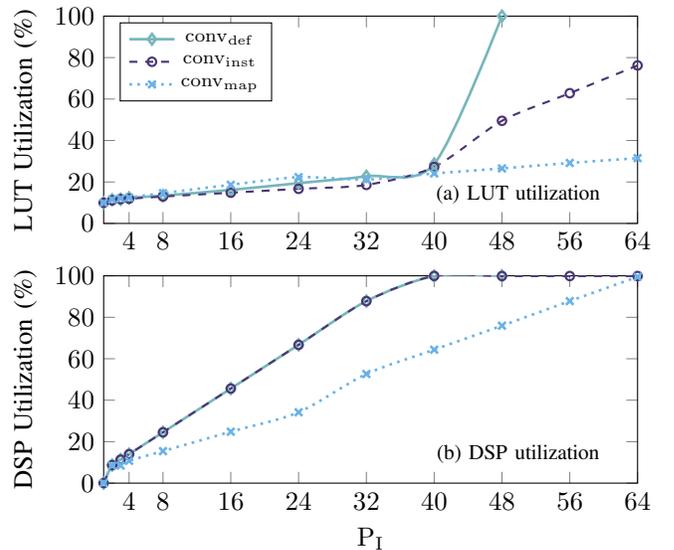
\begin{figure}[b]
    \begin{subfigure}[t]{0.98\columnwidth}
	\tikzsetnextfilename{conv1d_lut}

\begin{tikzpicture}
    \begin{axis}[
            xmin=1, 
            xmax=64, 
            ymin=0, 
            ymax=100, 
            xtick={4, 8, 16, 24, 32, 40, 48, 56, 64},
            ylabel={LUT Utilization (\%)},
            legend pos=north west,
            legend style={font=\scriptsize},
            width=\columnwidth,
            height=\columnwidth*0.5
            ] 

    \addplot [solid, smooth, mark=diamond,  mark options={solid}, color=RPTU_GreenGray, line width=1pt] table [x=instances, y=lut]{data/default_conv_utilization.txt}; 
    \addlegendentry{$\mathrm{conv}_\mathrm{def}$};

   \addplot [dashed, smooth, mark=o,  mark options={solid, scale=0.8}, color=RPTU_Violett, line width=0.8pt] table [x=instances, y=lut]{data/multi_inst_conv_utilization.txt}; 
    \addlegendentry{$\mathrm{conv}_\mathrm{inst}$};

    \addplot [dotted, smooth, mark=x,  mark options={solid}, color=RPTU_LightBlue, line width=1pt]  table [x=instances, y=lut]{data/multi_inst_conv_comb_utilization.txt}; 
    \addlegendentry{$\mathrm{conv}_\mathrm{map}$};

    \draw [black,-, thick] (axis cs:50,5) node[anchor=south] {\footnotesize (a) LUT utilization};
    
    \end{axis}

\end{tikzpicture}
     \centering\captionsetup{width=.8\linewidth}%
    \captionlistentry{}	
	\label{fig:conv1d_lut_util}
    \end{subfigure}\hspace{3mm}%
    \begin{subfigure}[t]{0.98\columnwidth}
	\tikzsetnextfilename{conv1d_dsp}

\begin{tikzpicture}
    \begin{axis}[
            xmin=1, 
            xmax=64, 
            ymin=0, 
            ymax=100, 
            xtick={4, 8, 16, 24, 32, 40, 48, 56, 64},
            xlabel={$\mathrm{P}_\mathrm{I}$},
            ylabel={DSP Utilization (\%)},
            legend pos=north west,
            legend style={font=\scriptsize},
            width=\columnwidth,
            height=\columnwidth*0.5
            ] 

    \addplot [solid, smooth, mark=diamond,  mark options={solid}, color=RPTU_GreenGray, line width=1pt] table [x=instances, y=dsp]{data/default_conv_utilization.txt}; 

   \addplot [dashed, smooth, mark=o,  mark options={solid, scale=0.8}, color=RPTU_Violett, line width=0.8pt] table [x=instances, y=dsp]{data/multi_inst_conv_utilization.txt}; 

    \addplot [dotted, smooth, mark=x,  mark options={solid}, color=RPTU_LightBlue, line width=1pt]  table [x=instances, y=dsp]{data/multi_inst_conv_comb_utilization.txt}; 

    \draw [black,-, thick] (axis cs:50,5) node[anchor=south] {\footnotesize (b) DSP utilization};

    \end{axis}

\end{tikzpicture}
    \centering\captionsetup{width=.8\linewidth}%
    \captionlistentry{}	
    \label{fig:conv1d_dsp_util}
    \end{subfigure}
    \caption{\ac{lut} and \ac{dsp} utilization of the different \ac{conv1d} layer architectures on the Xilinx \textit{XCVU13P}.}
    \label{fig:conv1d_util}
\end{figure}

\begin{figure*}[!t]
    \begin{subfigure}[c]{0.48\textwidth}
	\tikzsetnextfilename{SIM_35km_15dB_20GHz_TI1_II1_SL1024_SLPI128_LR02}
\def\colLight{60}

\begin{tikzpicture}
    \begin{axis}[
            ymode=log,
            log basis y={10},
            xmin=0, 
            xmax=0.608, 
            ymin=0.004, 
            ymax=1, 
            ylabel={BER},
            legend columns=4,
            legend style={at={(0.5,+0.75)},anchor=south, font=\scriptsize},
            width=\columnwidth,
            height=\columnwidth*0.5,
            restrict x to domain=0:0.608
            ] 

    \addplot [solid,  mark=none,  color=black, line width=1pt] table [x expr=\thisrow{Training_Time_Full_Throughput}*1e3, y=MEAN]{data/SIM_35km_15dB_20GHz_TI1_II1_SL1024_SLPI128_LR02.txt}; 
    \addlegendentry{Mean};
            
    \addplot[dashed, mark=none, RPTU_Violett, line width=1.0pt, domain=0:0.608] {0.027};
    \addlegendentry{FEC Threshold};
    
    
    \draw[line width=0.8pt, solid, color=RPTU_Orange] (0.320, 0.001) -- (0.320, 0.027);
    \addlegendimage{line width=0.8pt, solid, color=RPTU_Orange}
    \addlegendentry{$\mathrm{t}_\mathrm{conv}$};


    \begin{pgfonlayer}{background}

    \addplot [solid, mark=none, color=RPTU_GreenGray!\colLight, line width=0.4pt] table [x expr=\thisrow{Training_Time_Full_Throughput}*1e3, y=R1]{data/SIM_35km_15dB_20GHz_TI1_II1_SL1024_SLPI128_LR02.txt}; 

    \addplot [solid, mark=none, color=RPTU_GreenGray!\colLight, line width=0.4pt] table [x expr=\thisrow{Training_Time_Full_Throughput}*1e3, y=R2]{data/SIM_35km_15dB_20GHz_TI1_II1_SL1024_SLPI128_LR02.txt}; 

    \addplot [solid, mark=none, color=RPTU_GreenGray!\colLight, line width=0.4pt] table [x expr=\thisrow{Training_Time_Full_Throughput}*1e3, y=R3]{data/SIM_35km_15dB_20GHz_TI1_II1_SL1024_SLPI128_LR02.txt}; 

    \addplot [solid, mark=none, color=RPTU_GreenGray!\colLight, line width=0.4pt] table [x expr=\thisrow{Training_Time_Full_Throughput}*1e3, y=R4]{data/SIM_35km_15dB_20GHz_TI1_II1_SL1024_SLPI128_LR02.txt}; 

    \addplot [solid, mark=none, color=RPTU_GreenGray!\colLight, line width=0.4pt] table [x expr=\thisrow{Training_Time_Full_Throughput}*1e3, y=R5]{data/SIM_35km_15dB_20GHz_TI1_II1_SL1024_SLPI128_LR02.txt}; 

    \addplot [solid, mark=none, color=RPTU_GreenGray!\colLight, line width=0.4pt] table [x expr=\thisrow{Training_Time_Full_Throughput}*1e3, y=R6]{data/SIM_35km_15dB_20GHz_TI1_II1_SL1024_SLPI128_LR02.txt}; 

    \addplot [solid, mark=none, color=RPTU_GreenGray!\colLight, line width=0.4pt] table [x expr=\thisrow{Training_Time_Full_Throughput}*1e3, y=R7]{data/SIM_35km_15dB_20GHz_TI1_II1_SL1024_SLPI128_LR02.txt}; 

    \addplot [solid, mark=none, color=RPTU_GreenGray!\colLight, line width=0.4pt] table [x expr=\thisrow{Training_Time_Full_Throughput}*1e3, y=R8]{data/SIM_35km_15dB_20GHz_TI1_II1_SL1024_SLPI128_LR02.txt}; 

    \addplot [solid, mark=none, color=RPTU_GreenGray!\colLight, line width=0.4pt] table [x expr=\thisrow{Training_Time_Full_Throughput}*1e3, y=R9]{data/SIM_35km_15dB_20GHz_TI1_II1_SL1024_SLPI128_LR02.txt}; 

    \addplot [solid, mark=none, color=RPTU_GreenGray!\colLight, line width=0.4pt] table [x expr=\thisrow{Training_Time_Full_Throughput}*1e3, y=R10]{data/SIM_35km_15dB_20GHz_TI1_II1_SL1024_SLPI128_LR02.txt};

    \draw [black,-, thick] (axis cs:0.44,0.07) node[anchor=south] {\footnotesize (a) $\mathrm{P}_\mathrm{I}$: 33, $\mathrm{P}_\mathrm{T}$: \num{1}, SL: \num{512}};

    \end{pgfonlayer}

    \end{axis}

\end{tikzpicture}%
    \captionlistentry{}
	\label{fig:SIM_35km_15dB_20GHz_convergence_a}
    \end{subfigure}%
    \begin{subfigure}[c]{0.48\textwidth}
	\tikzsetnextfilename{SIM_35km_15dB_20GHz_TI1_II1_SL512_SLPI256_LR04}
\def\colLight{60}

\begin{tikzpicture}
    \begin{axis}[
            ymode=log,
            log basis y={10},
            xmin=0, 
            xmax=0.608, 
            ymin=0.004, 
            ymax=1, 
            ylabel={BER},
            legend pos=north west,
            legend style={font=\scriptsize},
            width=\columnwidth,
            height=\columnwidth*0.5,
            restrict x to domain=0:0.608
            ] 

    \addplot [solid,  mark=none,  color=black, line width=1pt] table [x expr=\thisrow{Training_Time_Full_Throughput}*1e3, y=MEAN]{data/SIM_35km_15dB_20GHz_TI1_II1_SL512_SLPI256_LR04.txt}; 
            
    \addplot[dashed, mark=none, RPTU_Violett, line width=1.0pt, domain=0:0.608] {0.027};

    
        \begin{pgfonlayer}{background}

    \addplot [solid, mark=none, color=RPTU_GreenGray!\colLight, line width=0.4pt] table [x expr=\thisrow{Training_Time_Full_Throughput}*1e3, y=R1]{data/SIM_35km_15dB_20GHz_TI1_II1_SL512_SLPI256_LR04.txt}; 

    \addplot [solid, mark=none, color=RPTU_GreenGray!\colLight, line width=0.4pt] table [x expr=\thisrow{Training_Time_Full_Throughput}*1e3, y=R2]{data/SIM_35km_15dB_20GHz_TI1_II1_SL512_SLPI256_LR04.txt}; 

    \addplot [solid, mark=none, color=RPTU_GreenGray!\colLight, line width=0.4pt] table [x expr=\thisrow{Training_Time_Full_Throughput}*1e3, y=R3]{data/SIM_35km_15dB_20GHz_TI1_II1_SL512_SLPI256_LR04.txt}; 

    \addplot [solid, mark=none, color=RPTU_GreenGray!\colLight, line width=0.4pt] table [x expr=\thisrow{Training_Time_Full_Throughput}*1e3, y=R4]{data/SIM_35km_15dB_20GHz_TI1_II1_SL512_SLPI256_LR04.txt}; 

    \addplot [solid, mark=none, color=RPTU_GreenGray!\colLight, line width=0.4pt] table [x expr=\thisrow{Training_Time_Full_Throughput}*1e3, y=R5]{data/SIM_35km_15dB_20GHz_TI1_II1_SL512_SLPI256_LR04.txt}; 

    \addplot [solid, mark=none, color=RPTU_GreenGray!\colLight, line width=0.4pt] table [x expr=\thisrow{Training_Time_Full_Throughput}*1e3, y=R6]{data/SIM_35km_15dB_20GHz_TI1_II1_SL512_SLPI256_LR04.txt}; 

    \addplot [solid, mark=none, color=RPTU_GreenGray!\colLight, line width=0.4pt] table [x expr=\thisrow{Training_Time_Full_Throughput}*1e3, y=R7]{data/SIM_35km_15dB_20GHz_TI1_II1_SL512_SLPI256_LR04.txt}; 

    \addplot [solid, mark=none, color=RPTU_GreenGray!\colLight, line width=0.4pt] table [x expr=\thisrow{Training_Time_Full_Throughput}*1e3, y=R8]{data/SIM_35km_15dB_20GHz_TI1_II1_SL512_SLPI256_LR04.txt}; 

    \addplot [solid, mark=none, color=RPTU_GreenGray!\colLight, line width=0.4pt] table [x expr=\thisrow{Training_Time_Full_Throughput}*1e3, y=R9]{data/SIM_35km_15dB_20GHz_TI1_II1_SL512_SLPI256_LR04.txt}; 

    \addplot [solid, mark=none, color=RPTU_GreenGray!\colLight, line width=0.4pt] table [x expr=\thisrow{Training_Time_Full_Throughput}*1e3, y=R10]{data/SIM_35km_15dB_20GHz_TI1_II1_SL512_SLPI256_LR04.txt}; 

    \end{pgfonlayer}

    \draw [black,-, thick] (axis cs:0.44,0.24) node[anchor=south] {\footnotesize (b) $\mathrm{P}_\mathrm{I}$: \num{33}, $\mathrm{P}_\mathrm{T}$: \num{1}, SL: \num{256}};

    \end{axis}

\end{tikzpicture}%
    \captionlistentry{}
	\label{fig:SIM_35km_15dB_20GHz_convergence_b}
    \end{subfigure}
    \begin{subfigure}[c]{0.48\textwidth}
    \vspace{-0.3cm}
	\tikzsetnextfilename{SIM_35km_15dB_20GHz_TI2_II0_SL512_SLPI256_LR04}
\def\colLight{60}

\begin{tikzpicture}
    \begin{axis}[
            ymode=log,
            log basis y={10},
            xmin=0, 
            xmax=0.608, 
            ymin=0.004, 
            ymax=1, 
            xlabel={Training Time (\si{\milli \second})},
            ylabel={BER},
            legend pos=north west,
            legend style={font=\scriptsize},
            width=\columnwidth,
            height=\columnwidth*0.5,
            restrict x to domain=0:0.608
            ] 

    \addplot [solid,  mark=none,  color=black, line width=1pt] table [x expr=\thisrow{Training_Time_Full_Throughput}*1e3, y=MEAN]{data/SIM_35km_15dB_20GHz_TI2_II0_SL512_SLPI256_LR04.txt}; 
            
    \addplot[dashed, mark=none, RPTU_Violett, line width=1.0pt, domain=0:0.608] {0.027};

    \draw[line width=0.8pt, solid, color=RPTU_Orange] (0.192, 0.001) -- (0.192, 0.027);

    
        \begin{pgfonlayer}{background}

    \addplot [solid, mark=none, color=RPTU_GreenGray!\colLight, line width=0.4pt] table [x expr=\thisrow{Training_Time_Full_Throughput}*1e3, y=R1]{data/SIM_35km_15dB_20GHz_TI2_II0_SL512_SLPI256_LR04.txt}; 

    \addplot [solid, mark=none, color=RPTU_GreenGray!\colLight, line width=0.4pt] table [x expr=\thisrow{Training_Time_Full_Throughput}*1e3, y=R2]{data/SIM_35km_15dB_20GHz_TI2_II0_SL512_SLPI256_LR04.txt}; 

    \addplot [solid, mark=none, color=RPTU_GreenGray!\colLight, line width=0.4pt] table [x expr=\thisrow{Training_Time_Full_Throughput}*1e3, y=R3]{data/SIM_35km_15dB_20GHz_TI2_II0_SL512_SLPI256_LR04.txt}; 

    \addplot [solid, mark=none, color=RPTU_GreenGray!\colLight, line width=0.4pt] table [x expr=\thisrow{Training_Time_Full_Throughput}*1e3, y=R4]{data/SIM_35km_15dB_20GHz_TI2_II0_SL512_SLPI256_LR04.txt}; 

    \addplot [solid, mark=none, color=RPTU_GreenGray!\colLight, line width=0.4pt] table [x expr=\thisrow{Training_Time_Full_Throughput}*1e3, y=R5]{data/SIM_35km_15dB_20GHz_TI2_II0_SL512_SLPI256_LR04.txt}; 

    \addplot [solid, mark=none, color=RPTU_GreenGray!\colLight, line width=0.4pt] table [x expr=\thisrow{Training_Time_Full_Throughput}*1e3, y=R6]{data/SIM_35km_15dB_20GHz_TI2_II0_SL512_SLPI256_LR04.txt}; 

    \addplot [solid, mark=none, color=RPTU_GreenGray!\colLight, line width=0.4pt] table [x expr=\thisrow{Training_Time_Full_Throughput}*1e3, y=R7]{data/SIM_35km_15dB_20GHz_TI2_II0_SL512_SLPI256_LR04.txt}; 

    \addplot [solid, mark=none, color=RPTU_GreenGray!\colLight, line width=0.4pt] table [x expr=\thisrow{Training_Time_Full_Throughput}*1e3, y=R8]{data/SIM_35km_15dB_20GHz_TI2_II0_SL512_SLPI256_LR04.txt}; 

    \addplot [solid, mark=none, color=RPTU_GreenGray!\colLight, line width=0.4pt] table [x expr=\thisrow{Training_Time_Full_Throughput}*1e3, y=R9]{data/SIM_35km_15dB_20GHz_TI2_II0_SL512_SLPI256_LR04.txt}; 

    \addplot [solid, mark=none, color=RPTU_GreenGray!\colLight, line width=0.4pt] table [x expr=\thisrow{Training_Time_Full_Throughput}*1e3, y=R10]{data/SIM_35km_15dB_20GHz_TI2_II0_SL512_SLPI256_LR04.txt}; 

\end{pgfonlayer}
    
    \draw [black,-, thick] (axis cs:0.44,0.24) node[anchor=south] {\footnotesize (c) $\mathrm{P}_\mathrm{I}$: \num{32}, $\mathrm{P}_\mathrm{T}$: \num{2}, SL: \num{256}};
    
    \end{axis}

\end{tikzpicture}%
    \captionlistentry{}
	\label{fig:SIM_35km_15dB_20GHz_convergence_c}
    \end{subfigure}%
    \begin{subfigure}[c]{0.48\textwidth}
    \vspace{-0.3cm}
	\tikzsetnextfilename{SIM_35km_15dB_20GHz_TI4_II0_SL1024_SLPI256_LR04}
\def\colLight{60}

\begin{tikzpicture}
    \begin{axis}[
            ymode=log,
            log basis y={10},
            xmin=0, 
            xmax=0.608, 
            ymin=0.004, 
            ymax=1, 
            xlabel={Training Time (\si{\milli \second})},
            ylabel={BER},
            legend pos=north west,
            width=\columnwidth,
            legend style={font=\scriptsize},
            height=\columnwidth*0.5,
            restrict x to domain=0:0.608
            ] 

    \addplot [solid,  mark=none,  color=black, line width=1pt] table [x expr=\thisrow{Training_Time_Full_Throughput}*1e3, y=MEAN]{data/SIM_35km_15dB_20GHz_TI4_II0_SL1024_SLPI256_LR04.txt}; 
            
    \addplot[dashed, mark=none, RPTU_Violett, line width=1.0pt, domain=0:0.608] {0.027};

    \draw[line width=0pt, solid, color=RPTU_Orange, opacity=0] (0.264, 0.001) -- (0.264, 0.027);

    
        \begin{pgfonlayer}{background}

    \addplot [solid, mark=none, color=RPTU_GreenGray!\colLight, line width=0.4pt] table [x expr=\thisrow{Training_Time_Full_Throughput}*1e3, y=R1]{data/SIM_35km_15dB_20GHz_TI4_II0_SL1024_SLPI256_LR04.txt}; 

    \addplot [solid, mark=none, color=RPTU_GreenGray!\colLight, line width=0.4pt] table [x expr=\thisrow{Training_Time_Full_Throughput}*1e3, y=R2]{data/SIM_35km_15dB_20GHz_TI4_II0_SL1024_SLPI256_LR04.txt}; 

    \addplot [solid, mark=none, color=RPTU_GreenGray!\colLight, line width=0.4pt] table [x expr=\thisrow{Training_Time_Full_Throughput}*1e3, y=R3]{data/SIM_35km_15dB_20GHz_TI4_II0_SL1024_SLPI256_LR04.txt}; 

    \addplot [solid, mark=none, color=RPTU_GreenGray!\colLight, line width=0.4pt] table [x expr=\thisrow{Training_Time_Full_Throughput}*1e3, y=R4]{data/SIM_35km_15dB_20GHz_TI4_II0_SL1024_SLPI256_LR04.txt}; 

    \addplot [solid, mark=none, color=RPTU_GreenGray!\colLight, line width=0.4pt] table [x expr=\thisrow{Training_Time_Full_Throughput}*1e3, y=R5]{data/SIM_35km_15dB_20GHz_TI4_II0_SL1024_SLPI256_LR04.txt}; 

    \addplot [solid, mark=none, color=RPTU_GreenGray!\colLight, line width=0.4pt] table [x expr=\thisrow{Training_Time_Full_Throughput}*1e3, y=R6]{data/SIM_35km_15dB_20GHz_TI4_II0_SL1024_SLPI256_LR04.txt}; 

    \addplot [solid, mark=none, color=RPTU_GreenGray!\colLight, line width=0.4pt] table [x expr=\thisrow{Training_Time_Full_Throughput}*1e3, y=R7]{data/SIM_35km_15dB_20GHz_TI4_II0_SL1024_SLPI256_LR04.txt}; 

    \addplot [solid, mark=none, color=RPTU_GreenGray!\colLight, line width=0.4pt] table [x expr=\thisrow{Training_Time_Full_Throughput}*1e3, y=R8]{data/SIM_35km_15dB_20GHz_TI4_II0_SL1024_SLPI256_LR04.txt}; 

    \addplot [solid, mark=none, color=RPTU_GreenGray!\colLight, line width=0.4pt] table [x expr=\thisrow{Training_Time_Full_Throughput}*1e3, y=R9]{data/SIM_35km_15dB_20GHz_TI4_II0_SL1024_SLPI256_LR04.txt}; 

    \addplot [solid, mark=none, color=RPTU_GreenGray!\colLight, line width=0.4pt] table [x expr=\thisrow{Training_Time_Full_Throughput}*1e3, y=R10]{data/SIM_35km_15dB_20GHz_TI4_II0_SL1024_SLPI256_LR04.txt}; 

\end{pgfonlayer}

    \draw [black,-, thick] (axis cs:0.44,0.24) node[anchor=south] {\footnotesize (d) $\mathrm{P}_\mathrm{I}$: \num{30}, $\mathrm{P}_\mathrm{T}$: \num{4}, SL: \num{256}};

    \end{axis}

\end{tikzpicture}%
    \captionlistentry{}
	\label{fig:SIM_35km_15dB_20GHz_convergence_d}
    \end{subfigure}%
    
    \caption{Convergence behavior of different hardware configurations for the optical fiber channel. Each thin line corresponds to a training run and the thick black line gives the mean of all training runs. The mean convergence time is given by the vertical orange line.} 
    \label{fig:SIM_35km_15dB_20GHz_convergence}
\end{figure*}
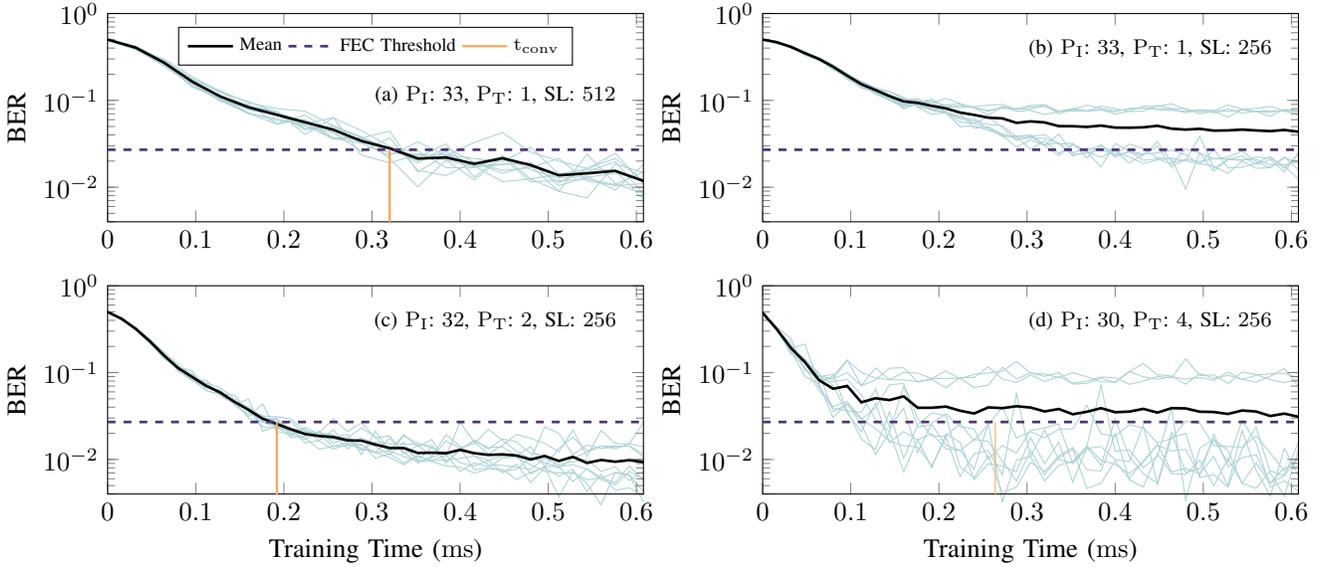

In Fig. \ref{fig:conv1d_util}, the utilization of \ac{lut} and \ac{dsp} resources after synthesis is shown for the different architectures. Since our custom mapping scheme can only be exploited by the inference module, we set $\mathrm{P}_\mathrm{T}$ to \num{1} and increase $\mathrm{P}_\mathrm{I}$ until the resources are fully utilized.  It can be seen that there is only a minor difference in the \ac{lut} usage up to $\mathrm{P}_\mathrm{I}=40$. Starting from $\mathrm{P}_\mathrm{I}=48$, the \ac{lut} resource consumption of $\mathrm{conv}_\mathrm{def}$ and $\mathrm{conv}_\mathrm{inst}$ increase significantly, while it grows only slightly for $\mathrm{conv}_\mathrm{map}$. The reason for this becomes clear when considering the \ac{dsp} utilization in Fig. \ref{fig:conv1d_dsp_util}. The \ac{dsp} utilization of $\mathrm{conv}_\mathrm{def}$ and $\mathrm{conv}_\mathrm{inst}$ increases linearly with $\mathrm{P}_\mathrm{I}$ with a similar slope. In contrast, the slope of $\mathrm{conv}_\mathrm{map}$ is much smaller since in this case the multiplications of two instances are mapped to a single \ac{dsp}. Therefore, the \ac{dsp} resources of $\mathrm{conv}_\mathrm{def}$ and $\mathrm{conv}_\mathrm{inst}$ are already fully utilized starting from $\mathrm{P}_\mathrm{I}=40$. For that reason, Vivado maps the remaining multiplications to \acp{lut}, which explains the sharp increase seen in Fig. \ref{fig:conv1d_lut_util}. 
In summary, the analysis shows that due to our optimized \ac{conv1d} layer architecture and our custom mapping scheme, the available \ac{dsp} resources can be used much more efficiently. This leads to a lower \ac{lut} utilization, a high number of implementable instances, and eventually increases the throughput achievable by our \ac{cnn} hardware implementation. 

\subsection{Convergence Behavior}
\label{sec:convergence_behavior}
In this section, the convergence behavior of our \ac{cnn}-based equalizer is analyzed for the optical channel described in Sec. \ref{sec:channel_model}. To evaluate the performance and convergence behavior of our hardware architecture, we train different hardware configurations for the described channel model. The main goal of this analysis is to investigate the influence of the sequence length (training symbols per weight-update) and the number of training instances on the convergence time for an unseen channel. The different configurations are implemented on the Xilinx XCVU13P with a clock frequency of \SI{150}{\mega \hertz}, which is the highest frequency achieving timing closure. To satisfy the throughput requirement of \SI{20}{\giga Bd} with this clock frequency, \num{34} parallel \ac{cnn} instances are required, thus $\mathrm{P}_\mathrm{I} + \mathrm{P}_\mathrm{T} \geq 34$. 

In Fig. \ref{fig:SIM_35km_15dB_20GHz_convergence} the result of this analysis is shown. SL gives the length of each training sequence, corresponding to the number of symbols each training instance processes for one weight-update. The learning rate is set to \num{0.001} for all configurations since a higher learning rate results in unstable behavior and a lower one increases the training time. We define the convergence time $\mathrm{t}_\mathrm{conv}$ as the training time required to achieve a \ac{ber} below the \ac{fec} threshold of \num{2.7e-2} which corresponds to a \ac{ldpc} code with \SI{20}{\percent} overhead~\cite{Agrell2018}. 
When comparing Fig. \ref{fig:SIM_35km_15dB_20GHz_convergence_a} and \ref{fig:SIM_35km_15dB_20GHz_convergence_b}, it can be seen how the sequence length influences the convergence behavior. For the configuration with the largest sequence length of \num{512} (Fig. \ref{fig:SIM_35km_15dB_20GHz_convergence_a}), the models of all training runs eventually converge to the \ac{fec} threshold, with a mean convergence time of \SI{0.264}{\milli \second}. With a lower sequence length, the convergence behavior becomes more unstable resulting in only \SI{40}{\percent} of the models converging for a sequence length of \num{256}. One way to reduce the convergence time while maintaining stable convergence is to increase the number of training instances $\mathrm{P}_\mathrm{T}$. This way, the weights are updated multiple times based on the gradients of multiple independent sub-sequences. The results of using a different number of training instances can be seen when comparing Fig. \ref{fig:SIM_35km_15dB_20GHz_convergence_b}, \ref{fig:SIM_35km_15dB_20GHz_convergence_c}, and \ref{fig:SIM_35km_15dB_20GHz_convergence_d}. 
When using two training instances, as shown in Fig. \ref{fig:SIM_35km_15dB_20GHz_convergence_c}, each training instance performs a weight-update based on the same initial weights. Therefore, a similar behavior as for the configuration with one training instance  (Fig. \ref{fig:SIM_35km_15dB_20GHz_convergence_a}) can be observed although the sequence length is much smaller. However, further increasing $\mathrm{P}_\mathrm{T}$ results in unstable convergence, as shown in Fig. \ref{fig:SIM_35km_15dB_20GHz_convergence_d}. Due to the parallel structure of our hardware architecture, the weight-updates of all training instances are based on the same initial weights. Thus, each training instance updates the weights with a gradient pointing in a similar direction, resulting in a larger step size and unstable convergence. Experiments showed that even a lower learning rate can't compensate for this effect.  
To summarize, the fastest stable convergence is achieved by the configuration of Fig. \ref{fig:SIM_35km_15dB_20GHz_convergence_c} with two training instances and a sequence length of \num{256}, resulting in a convergence time of \SI{0.144}{\milli \second}.

\subsection{Comparison to Embedded GPU}
\begin{table}[t]
\centering
\caption{Hardware implementation results}
\label{tab:convergence_implementation_results}
\begin{tabular}{ccccc|cc}

\toprule
 \multirow{2}{*}{Platform}  & \multirow{2}{*}{$\mathrm{P}_\mathrm{I}$}           & \multirow{2}{*}{$\mathrm{P}_\mathrm{T}$}      & TP        &  $\mathrm{t}_\mathrm{conv}$   & LUT         & DSP    \\
 &       &      & (\si{\giga Bd}) & (\si{\milli \second}) &(\si{\percent})  & (\si{\percent})\\

& & & & & & \\[-1.8ex]
\cline{1-7} 
& & & & & & \\[-1.8ex]

\multirow{2}{*}{FPGA} & \num{33} &  \num{1} & \num{20} & \num{0.320} & \num{25.18} & \num{54.87}  \\
& \num{32} &  \num{2} & \num{20} & \num{0.192} & \num{40.47} & \num{57.46} \\

& & & & & & \\[-2.4ex]
\cline{1-7} 
& & & & & & \\[-1.8ex]

\multirow{4}{*}{GPU} &\num{34} &  \num{0} & \num{2.6e-2} & --- &\multirow{4}{*}{\shortstack[c]{\ac{cnn}\\mode:}} & \multicolumn{1}{l}{Inference} \\
& \num{0} &  \num{34} & \num{4.7e-3} & \num{910} &  &  \multicolumn{1}{l}{Training}   \\
& \num{8192} &  \num{0} & \num{7.4e-1} & --- & &  \multicolumn{1}{l}{Inference} \\
& \num{0} &  \num{8192} & \num{1.9e-1} & \num{3500} & & \multicolumn{1}{l}{Training}  \\

\bottomrule
\end{tabular}
\end{table}
In Tab. \ref{tab:convergence_implementation_results}, we give the \ac{fpga} implementation results of the two configurations achieving stable convergence and compare them to the same \ac{cnn} running on the embedded \ac{gpu} Jetson AGX Xavier. For the \ac{gpu}, it is not possible to process the \ac{cnn} simultaneously in training and inference mode while keeping the weights synchronized. Thus, the model is either set to inference mode or to training mode, where $\mathrm{P}_\mathrm{I}$ corresponds to the inference batch size and $\mathrm{P}_\mathrm{T}$ corresponds to the training batch size. In Tab. \ref{tab:convergence_implementation_results} it can be seen that increasing $\mathrm{P}_\mathrm{T}$ from \num{1} to \num{2} for the \ac{fpga}, increases the \ac{lut} utilization by \SI{15}{\percent} which shows that the computational complexity of the training module is much higher as compared to the inference module. This validates our approach of using a distinct inference module with high parallelism since satisfying the throughput requirements with training modules only would not be feasible. Further, the high throughput is enabled by our custom \ac{dsp} mapping. With the other \ac{conv1d} architectures of Sec. \ref{sec:hardware_architecture_analysis} the design becomes unrouteable due to high utilization and congestion.

For the \ac{gpu} we set the batch size to \num{34}, which is the same as for the \ac{fpga}, and to \num{8192} which is the batch size where the throughput of the \ac{gpu} begins to saturate. It can be seen that even if the model runs in inference mode, without including training, the highest achievable throughput is only \SI{0.19}{\giga Bd}. This is two orders of magnitude lower than that of the \ac{fpga}, which indicates that it is not feasible to satisfy our throughput requirements of \SI{20}{\giga Bd} with conventional platforms. 

Due to the lower throughput, the convergence time $\mathrm{t}_\mathrm{conv}$ is orders of magnitude higher than that of the \ac{fpga}. It lies in the order of seconds, which is too high for most practical communication scenarios. For instance, \ac{pon} requires a reconfiguration capacity in the order of \si{\micro \second} \cite{Simon2020},  while a \SI{2}{\giga \hertz} wireless channel with a receiver velocity of \SI{108}{\kilo \meter \per \hour} has a coherence time of \SI{2.5}{\milli \second} \cite{Morocho2020}.

In summary, it can be seen that our custom \ac{fpga} architecture is much better suited for the application of high-throughut \ac{cnn}-based equalization than a general purpose \ac{gpu}. This can be attributed to the fact that \acp{gpu} are well suited for the processing of large  \acp{nn}. In contrast, the computational resources can't be utilized efficiently for \acp{nn} of low complexity. In this case, a customized hardware architecture provides much better results.


\subsection{Comparison to State-of-the-Art}

\begin{table}[t]
\centering
\caption{Comparision to state-of-the-art}
\label{tab:comparision_to_state_of_the_art}
\begin{tabular}{ccccccc}

\toprule
& \cite{Tang2022} & \cite{Venkataramanaiah2019} & \cite{Liu2017} & \cite{Hong2021} & \cite{Ney2022} & Ours \\
\midrule

$\mathrm{f}_\mathrm{clk}$ (\si{\mega \hertz}) & \num{100} & \num{240} & \num{200} & \num{100} & \num{300} & \num{150} \\
Batch Size & 128 & 40 & 16 & 1 & 1 & 34 \\
GOPS & \num{28.15} & \cellcolor{RPTU_LightGreen!30} \num{479} & \num{86.12} & \num{4.39} &  \num{5.28} & \num{236} \\
TP (\si{\mega pixel \per \second}) & \num{0.5} & \num{3.4} & --- & \num{30} &  \num{49} & \cellcolor{RPTU_LightGreen!30} \num{20000} \\
Power (\si{\watt}) & \num{6.89} & \num{50.47} & \num{14.24} & \num{0.67} & \cellcolor{RPTU_LightGreen!30} \num{0.57} & \num{22.34} \\
$\eta$ (GOPS/W) & \num{4.09} & \num{9.49} & \num{6.05} & \num{6.55} & \num{9.66} & \cellcolor{RPTU_LightGreen!30} \num{10.56}\\

\bottomrule
\end{tabular}

\end{table}

In this section, our hardware architecture is compared to other state-of-the-art \ac{nn}-training \ac{fpga} implementations introduced in Sec. \ref{sec:nn_training_on_fpga}. 
Since there are large variations in the different works (\ac{nn} topology, target platform, requirements, etc.), for a fair comparison, we mainly focus on the \ac{gops} and on the efficiency $\eta$ given as \si{GOPS \per \watt}. Further, we give the throughput in terms of \si{\mega pixel \per \second} as a measurement of how many input pixels can be processed per time unit. This allows us to compare the throughput of our 1D-\ac{cnn} to the two-dimensional \ac{cnn} architectures of the related works and to the throughput requirements of our channel. 

In Tab. \ref{tab:comparision_to_state_of_the_art}, it can be seen that the highest \ac{gops} is achieved by \cite{Venkataramanaiah2019} with \num{479} while our work provides \num{236} \ac{gops}.
However, higher \ac{gops} does not necessarily result in a higher throughput in terms of \si{pixel \per \second}, since there is a huge difference in the \ac{nn} topology of the different works. With respect to throughput, our work provides the best results with \SI{20}{\giga pixel \per \second}, corresponding to \SI{20}{\giga Bd} for our application. Achieving this throughput is essential to satisfy the communication requirements of our communication channel. This is mainly achieved due to our low-complex \ac{cnn} topology in combination with our efficient hardware architecture which enables the high inference parallelism. In contrast, the related works mostly focus on the training of much larger \ac{cnn} topologies making it unfeasible to meet the high throughput requirements. 

In terms of efficiency, our work achieves the best result with \SI{10.56}{GOPS \per \watt}. This can be explained by optimized hardware architecture and our efficient mapping of multiplications to \ac{dsp} resources. 

In summary, the huge variety of different works makes it challenging to provide a fair comparison, however, it becomes clear that our work achieves comparable performance as state-of-the-art hardware accelerators. In particular, for the application of high-throughput equalization, our work fills a gap regarding trainable \ac{nn} \ac{fpga} implementations.

\section{Conclusion}
\label{sec:conclusion}

In this work, we present an optimized high-throughput \ac{fpga} implementation of a \ac{nn}-based equalizer for communications. The hardware architecture exploits batch-level parallelism in the convolutional layer to enable a custom mapping scheme of two multiplications to a single \ac{dsp} and provides a variable \ac{dop} for the \ac{nn} inference and the training. The performance of our implementation is demonstrated for an optical fiber \ac{imdd} channel with a data rate of \SI{20}{\giga Bd}. It is shown that our approach satisfies the high throughput requirements while allowing for adaptation to different channel conditions by training the \ac{nn} on the \ac{fpga} which is validated by a detailed convergence analysis. Further, results show, that our \ac{fpga} implementation outperforms an embedded \ac{gpu} in terms of throughput and provides state-of-the-art performance in terms of energy efficiency as compared to other \ac{nn} training \ac{fpga} implementations. To summarize, our work provides a foundation towards \ac{cnn}-based equalization on \ac{fpga}, which is a crucial step for the development of flexible, high-throughput, next-gen communication systems.

\section*{Acknowledgements}
\label{sec:acknowledgements}

\ifthenelse{\boolean{blind}}
{
\textit{Acknowledgementss omitted due to double-blind review}
}
{
We sincerely thank Prof. Laurent Schmalen and Vincent Lauinger from the Communications Engineering Lab (CEL) of Karlsruhe Institute of Technology (KIT) for providing the channel model and for the insightful discussions. 
}

\bibliographystyle{IEEEtran}
\bibliography{IEEEabrv, bib_bibtex}

\end{document}